\documentclass[12pt, a4paper]{article}

\usepackage[T1]{fontenc}
\usepackage [english]{babel}
\usepackage{natbib} 
\usepackage{authblk}
\usepackage [english]{babel}
\usepackage{booktabs}
\usepackage{adjustbox}
\usepackage[table,xcdraw]{xcolor}
\usepackage[hmargin=2.5cm,vmargin=2cm]{geometry}
\usepackage{listings}
\usepackage{xcolor}
\usepackage{amsmath}
\usepackage{dsfont}
\usepackage{adjustbox}
\usepackage{amssymb}
\usepackage{amsfonts}
\usepackage{bm}
\usepackage{amsthm} 
\usepackage{graphicx}
\usepackage{enumitem}
\usepackage{fontspec}
\defaultfontfeatures{Extension = .otf}
\usepackage{fontawesome} 
\usepackage[colorlinks=true,citecolor=blue,linkcolor = blue,urlcolor  = blue]{hyperref}
\usepackage{soul}
\sethlcolor{hl}
\usepackage[many]{tcolorbox}
\usepackage{natbib}
\usepackage{ragged2e} 
\usepackage{threeparttable}
\usepackage{float}
\floatstyle{plaintop}
\restylefloat{table}
\justifying
\usepackage[linesnumbered,lined,ruled,commentsnumbered]{algorithm2e}
\usepackage{setspace}

\def\vector#1{\mbox{\boldmath{$#1$}}} 

\newcommand{\Xit}{\mathbf{X}_{it}}
\newcommand{\Dit}{D_{it}}
\newcommand{\Yit}{Y_{it}}

\newcommand{\Uit}{U_{it}}

\newcommand{\Vit}{V_{it}}

\newcommand{\bmeta}{\bm{\eta}}

\newcommand{\E}{\mathbb{E}}

\newcommand{\indep}[3]{#1 \perp\kern-5pt \perp #2 \mid #3}

\newcommand{\one}{\mathds{1}}

\newcommand{\Wk}{\mathcal{W}_k}

\newtheorem{assumption}{Assumption}

\newtheorem{remark}{Remark}

\renewcommand{\theequation}{\arabic{section}.\arabic{equation}}

\usepackage{authblk}
\title{\Large{Double Machine Learning with High-dimensional Interactive Fixed Effects}}

\author[$^{\dagger}$]{Binzhi Chen}
\author[$^{\dagger}$]{Annalivia Polselli}
\author[$^{\dagger}$]{ Paul S. Clarke\thanks{
\textbf{Authors' contact details:} Binzhi Chen: \url{Binzhi.Chen@essex.ac.uk}. Annalivia Polselli: \url{annalivia.polselli@essex.ac.uk}. Paul S. Clarke: \url{pclarke@essex.ac.uk}. Binzhi Chen and Paul S. Clarke acknowledge funding from the UK Economic and Social Research Council award ES/S012486/1 (MiSoC). Annalivia Polselli acknowledges support of the British Academy through the Postdoctoral Fellowship (grant number PFSS24/240003). 
The authors acknowledge the use of the High Performance Computing Facility (Ceres) and its associated support services at the University of Essex in the completion of this work.
}}
\affil[$^{\dagger}$]{\footnotesize Institute for Social and Economic Research, University of Essex, Colchester CO4 3SQ, UK}
\date{Preliminary draft: \today}

\begin{document}
\maketitle

\begin{abstract}
\setstretch{1.1}
Factor structures are central to empirical work in economics and finance, and are usually used to model time-varying unobserved heterogeneity through interactive fixed effects (IFE). Existing IFE estimators rest on low-dimensional and linear specifications in the covariates, assumptions which are increasingly restrictive in applications drawing on rich datasets with controls of unknown functional form. This paper develops a Double Machine Learning estimator for the high-dimensional partially linear panel model with interactive fixed effects (panel DML-IFE). The method combines projection-based defactorisation of the data, in the spirit of Common Correlated Effects (CCE), with a Neyman-orthogonal score function and cross-fitting procedure, and accommodates low-rank factor structures in outcomes and treatments alongside high-dimensional, potentially nonlinear covariate effects estimated by machine learning algorithms. Monte Carlo simulations show that panel DML-IFE outperforms conventional IFE estimator outside the correctly-specified linear case, with bias reduction driven primarily by the time and covariate dimensions. An empirical application to U.S. stock returns shows that several effects documented under linear specifications lose statistical significance once high-dimensional nonlinear confounding and the presence of IFE are jointly accounted for.

\medskip
\noindent \textbf{Keywords:} Base learners, causal machine learning, common correlated effects, Neyman Orthogonality, panel data. \\
\textbf{JEL codes:}  C14, C18, C33, C45, C52, C58.
\end{abstract}

\newpage 

\section{Introduction}
Panel data models with interactive fixed effects (IFE), formalised by \citet{pesaran2006estimation} and \citet{bai2009panel}, are central to the identification of structural parameters in the presence of time-varying unobserved heterogeneity, which makes the standard within-group estimator inconsistent. IFE structures arise naturally in asset pricing and risk analysis, where returns are driven by a small number of latent risk factors with heterogeneous loadings across assets, as formalised in the Arbitrage Pricing Theory \citep{ross1976arbitrage}, the Fama-French three-factor model \citep{fama1993common}, in large-scale portfolio performance frameworks \citep{connor1986performance}, and in high-dimensional approximate factor models \citep{chamberlain1982arbitrage}. They are equally central in macroeconometrics, where latent factors capture common drivers of economic activity and inflation \citep{stock2002macroeconomic}, and in consumer theory, where they represent unobservable taste shocks \citep{nevo2001measuring}.

The bulk of the econometric literature on IFE rests on two restrictive assumptions: the set of observed covariates $p$ is small relative to the panel dimensions $(N,T)$, and these covariates enter the estimating equation linearly \citep[see, e.g.,][]{ando2016panel,li2016panel,moon2017dynamic,moon2018nuclear,miao2020panel,karavias2023structural,ditzen2025interactive,chen2025iv}. Both assumptions are increasingly hard to defend in empirical work, where applications in finance, labour, and trade routinely draw on rich administrative and financial datasets in which the conditioning set is high-dimensional and of unknown functional form. \citet{rucker2025} is a notable exception, incorporating a desparsified Lasso into the Common Correlated Effects (CCE) framework of \citet{pesaran2006estimation} to handle a high-dimensional conditioning set of covariates, but the broader IFE literature has not engaged yet with the wider class of machine learning estimators (e.g., tree-based methods and neural networks) or with the inferential guarantees they require. The Double Machine Learning (DML) framework of \citet{chernozhukov2018double} establishes precisely these guarantees, delivering valid inference on low-dimensional structural parameters in the presence of high-dimensional nuisance components estimated by any machine learning algorithm. However, no existing DML framework accommodates the interactive fixed-effects structure that arises naturally in panels with latent common shocks and heterogeneous loadings.

We develop Double Machine Learning estimation for the high-dimensional partially linear panel regression model with interactive fixed effects, henceforth \emph{panel DML-IFE}. The procedure combines the DML framework of \citet{chernozhukov2018double} with the high-dimensional CCE approach of \citet{rucker2025} to accommodate the use of \emph{any} machine learning algorithm in the prediction phase rather than just desparsified Lasso;\ this last feature is important for robustness because it allows the use of ensemble learning strategies that choose the best-performing learning algorithm and thus minimise bias.  The outcome and treatment equations admit low-rank factor structures, and the observed covariates may themselves exhibit high-dimensional factor structures that enter the outcome and treatment through unknown nuisance functions.  
The central methodological challenge is to remove the interactive fixed effects by integrating projection-based defactorisation into an orthogonalised DML procedure that delivers valid causal inference under a general class of learners converging at the standard $(NT)^{-1/4}$ rate, without solving non-convex low-rank matrix problems. We address it by adopting the projection-based defactorisation of \citet{pesaran2006estimation}, following \citet{rucker2025}, which approximates the factor space using cross-sectional averages of the observables and avoids the non-convex optimisation underlying PCA-based alternatives. Applying the Neyman-orthogonal score to the defactorised variables and cross-fitting at the unit level yields a $\sqrt{NT}$-consistent and asymptotically normal estimator of the structural parameter, with a cluster-robust variance estimator that delivers valid confidence intervals under standard DML rate conditions. The resulting framework includes the within-group estimator, two-way fixed effects (TWFE), and grouped fixed effects (GFE) as special cases.

Our framework innovates by combining three strands of the literature. First, we extend the IFE panel data literature \citep[][]{chudik2011,chudik2015,ando2016panel,li2016panel,ando2017clustering,moon2017dynamic,moon2018nuclear,miao2020panel,juodis2022,gao2023binary,karavias2023structural,ditzen2025multiple,ditzen2025interactive,chen2025iv} to settings with high-dimensional covariate sets that may enter the outcome and treatment equations nonlinearly. While our framework is motivated by the factor structure of \citet{bai2009panel}, we deliberately avoid direct estimation via principal components due to its non-convexity and computational cost in high-dimensional settings.\footnote{We do not adopt the Principal Component Analysis (PCA) approach of \citet{bai2009panel}, whose non-convex low-rank optimisation becomes computationally prohibitive once flexible learners are layered on top, nor the convex relaxation through nuclear-norm penalisation of \citet{moon2018nuclear}, which remains computationally intensive and may be sensitive to regularisation choices.} 
Instead, we adopt the projection-based method of \citet{pesaran2006estimation}, which removes interactive fixed effects through cross-sectional averages and is particularly well suited to high-dimensional settings predicted with machine learning devices. The projection estimator typically achieves $\sqrt{NT}$-consistency (or a rate close to it) and so does not need to be treated as a DML nuisance parameter.  This combination of projection-based defactorisation with a any machine learning device in the prediction phase is, to our knowledge, the first integration of DML and IFE in the literature. The closest precedent is \citet{rucker2025}, who extend CCE to high-dimensional settings via a desparsified Lasso; they restrict attention to Lasso and to the associated theoretical guarantees, whereas we accommodate a general class of machine learning algorithms and deliver valid inference for it. 
 
Second, we extend the DML framework of \citet{chernozhukov2018double} to interactive fixed-effects panels. Existing DML developments for panel data \citep{klosin2022,semenova2023inference,clarke2025double,marquez2025, baiardi2026} assume additive separability of the unobserved heterogeneity.  We relax this assumption to allow the latent component to vary multiplicatively across units and time and thus extend the panel DML procedure proposed by \citet{clarke2025double}. Our approach formally extends the Neyman orthogonalisation and cross-fitting approach to a context where nuisance components consist not only of high-dimensional covariate functions, but also high-dimensional factor loadings estimated from the data. This allows consistent estimation of treatment effects even when the unobserved confounders evolve dynamically and affect both treatment assignment and outcomes.

Third, we contribute to the empirical literature that uses IFE to capture time-varying unobserved heterogeneity in panel applications. IFE structures have been widely used in cross-country and regional analyses to capture latent common shocks and heterogeneous responses across units \citep{gobillon2016regional,xu2017generalized,callaway2023treatment}. In development and migration research, accounting for unobserved multilateral resistance or global shocks is crucial to avoid omitted-variable bias and spurious inference \citep{bertoli2013multilateral}. Similarly, in macroeconomic and environmental applications, unobserved common factors and heterogeneous loadings play a central role in identifying causal relationships \citep{bolat2023there,rafaty2025carbon}.
The panel DML-IFE framework enables such applications to incorporate high-dimensional and flexibly adjusted controls without sacrificing valid inference, broadening the range of settings in which interactive fixed effects can be combined with credible identification of structural parameters.

We assess the finite sample performance of the panel DML-IFE estimator through Monte Carlo simulations across three nuisance-function designs (linear, nonlinear and smooth, nonlinear and discontinuous), under a range of various configurations calibrated to our empirical application. Panel DML-IFE outperforms the linear IFE estimator of \citet{bai2009panel} outside the correctly-specified linear design, recovers complex covariate dependence even when non-linear basis terms are not explicitly supplied to the learner, and exhibits bias attenuation primarily in time dimension rather than in the cross-sectional dimension, indicating that reliable inference is possible in macro panels and short surveys provided the time and covariate dimensions are sufficiently rich.

We illustrate the empirical relevance of the method in an application to the U.S. stock returns, building on the data and design of \citet{rucker2025}. Using a panel of $N=29$ large-cap Dow Jones constituents observed over $T=72$ monthly periods, with $p=89$ lagged firm characteristics, we estimate the effect of several candidate return predictors. 
The comparison between IFE and panel DML-IFE reveals two recurring patterns once nonlinear confounding and interactive fixed effects are jointly accounted for. First, panel DML-IFE recovers significant effects of inventory changes and of scaled earnings forecasts on excess returns that IFE fails to detect. The pattern is consistent with nonlinear confounding by high-dimensional firm characteristics, which the linear IFE specification cannot accommodate. Second, when the underlying relationship is approximately linear (as for turnover volatility) or when no genuine effect is present (as for the bid-ask spread) panel DML-IFE and IFE produce similar estimates, confirming that the added flexibility of panel DML-IFE neither artificially generates findings nor inverts signs.

The rest of the article is structured as follows. Section~\ref{sec:model} introduces the model and assumptions. 
Section~\ref{sec:algorithm} presents the estimation procedure. Section~\ref{sec:mcsims} reports the Monte Carlo evidence. Section~\ref{sec:empirical} presents the asset-pricing application. Section~\ref{sec:conclusion} concludes. 

\paragraph{Notation.} For $A\in\mathbb{R}^{m\times n}$, $\|A\|_{\mathrm{op}}=\sigma_{\max}(A)$ denotes the operator (spectral) norm and $\|A\|_F=\bigl(\sum_{i,j}A_{ij}^2\bigr)^{1/2}=\bigl(\sum_k \sigma_k(A)^2\bigr)^{1/2}$ the Frobenius norm; the two satisfy $\|A\|_{\mathrm{op}}\le\|A\|_F\le\sqrt{\mathrm{rank}(A)}\,\|A\|_{\mathrm{op}}$, together with the submultiplicativity $\|AB\|_{\mathrm{op}}\le\|A\|_{\mathrm{op}}\|B\|_{\mathrm{op}}$. We write $\mu_{\max}(A)$ for the largest eigenvalue of a square matrix $A$, $\mathrm{tr}(A)=\sum_{i=1}^N a_{ii}$ for its trace, $P_A=A(A'A)^{-1}A'$ for the projection onto the column space of $A'$, and $M_A=I-P_A$ for its orthogonal complement. We set $\delta_{NT}^2=\min\{N,T\}$ and write $(N,T)\to\infty$ for joint divergence. We write $a_T \asymp b_T$ if $a_T=O\left(b_T\right)$ and $b_T=O\left(a_T\right)$.

\section{Model and Assumptions}\label{sec:model}
\setcounter{equation}{0}
\setcounter{lemma}{0}
\setcounter{proposition}{0}

\subsection{The Partially Linear Panel Regression Model with IFE}
Suppose we observe a balanced panel dataset consisting of $N$ individuals over $T$ time periods.\footnote{The framework also accommodates panels in which the time periods are equally spaced but not consecutive, and unbalanced panels in which a unit's time series ends before the end of the sample.} For each individual $i = 1, \dots, N$ and each time period $t = 1, \dots, T$, we observe the outcome variable $Y_{it} \in \mathbb{R}$, a treatment variable $D_{it} \in \mathcal{D}$ (which may be binary, continuous, or categorical), and a $p$-dimensional vector of observed covariates $\mathbf{X}_{it} \in \mathbb{R}^p$, where $p$ can be greater than $N$ and $T$ to capture high-dimensional data scenarios. 

Using this notation, we propose the following partially linear panel regression model \citep{robinson1988root} with interactive fixed effects (henceforth, panel DML-IFE) to capture time-varying omitted confounding:
\begin{align}
{Y}_{it} &= \theta_0 V_{it} + l_0(\Xit) + \boldsymbol{\lambda}_i^\prime \mathbf{f}_t  +{U}_{it}, \label{eq:basic_model1} \\
{V}_{it} &= {D}_{it} - m_0(\Xit) - \boldsymbol{\phi}_i^{\prime}\mathbf{f}_t, \label{eq:basic_model2} \\
\Xit &= {\Gamma_i' \mathbf{f}_t} + \mathbf{E}_{it},
\label{eq:basic_model3}
\end{align}
where $\mathbb{E}[U_{it} \mid \Dit, \Xit, \mathbf{f}_t, \boldsymbol{\lambda}_i] = \mathbb{E}[V_{it} \mid \Xit, \mathbf{f}_t, \boldsymbol{\phi}_i] = 0$ and $\mathbb{E}[\mathbf{E}_{it} \mid \Xit, \mathbf{f}_t, \Gamma_i] = {\bf 0}$.\footnote{The framework accommodates binary outcomes $Y_{it}\in\{0,1\}$, in which case $\theta_0$ is an average partial effect on the probability scale and heteroskedasticity-robust standard errors are required because the residual variance depends on the fitted probability. In this case, inference remains valid but less efficient than under a correctly specified nonlinear link. Binary or multi-categorical treatments $D_{it}$ are accommodated analogously. The treatment equation is interpreted as a linear projection of $D_{it}$ onto $(\mathbf{X}_{it}, \phi_i, \mathbf{f}_t)$, with $V_{it}$ the associated projection residual, by construction. This is a modelling choice rather than a distributional assumption on $V_{it}$. Since identification of $\theta_0$ requires $\mathbb{E}[V_{it}\mid \mathbf{X}_{it}, \phi_i, \mathbf{f}_t] = 0$, then $m_0(\mathbf{X}_{it}) + \phi_i' \mathbf{f}_t$ coincides with $\Pr(D_{it}=1\mid \mathbf{X}_{it}, \phi_i, \mathbf{f}_t)$ and
therefore lies in $[0,1]$ almost surely. In practice, this can be enforced by trimming observations whose estimated propensity falls outside $[\tau, 1-\tau]$ for a small $\tau > 0$.}
The time-varying fixed effects are modelled as depending on $\mathbf{f}_t \in \mathbb{R}^r$ an unobserved $r$-dimensional vector of common latent factors at time $t$. Parameter $\boldsymbol{\lambda}_i \in \mathbb{R}^r$ represents the factor loadings  in $Y_{it}$,  $\boldsymbol{\phi}_i$ is the equivalent for the factor loadings in $V_{it}$ and  $\boldsymbol{\Gamma}_i$ is the equivalent for the factor loadings in $X_{it}$. 

\begin{remark}\label{rem:iv_approach}
The model is written using the `partialled out' approach of \citet{robinson1988root} with $V_{it}$ rather than $D_{it}$ representing the treatment. This is convenient for DML because it means that nuisance parameters $l_0(\cdot)$ and $m_0(\cdot)$ can be learnt directly from the transformed data (the transformation to be defined below). From \citet[C3-C4]{chernozhukov2018double}, the model specification of (\ref{eq:basic_model1})  using the alternative `IV approach', i.e., ${Y}_{it} = \theta_0 D_{it} + g_0(\Xit) + \boldsymbol{\lambda}_i^\prime \mathbf{f}_t  +{U}_{it}$, where $g_0(\cdot)\neq l_0(\cdot)$, is less attractive because model fitting generally requires iterating between learning $g_0(\cdot)$ and estimating $\theta_0$ until convergence.
\end{remark}

Following \cite{clarke2025double}, the effects of the observed and unobserved time-varying confounding are taken to be additively separable. 
Imposing additive separability ensures that the high-dimensional factor structure remains separable from the nonlinear covariates transformations, thereby maintaining the validity of the theoretical framework developed in this paper.

\begin{remark}
The model generalises standard fixed-effects panel specifications by allowing for a low-rank factor structure in the unobservables. Specifically, the terms $\Gamma_i'\mathbf{f}_t$, $\boldsymbol{\lambda}_i'\mathbf{f}_t$ and  $\boldsymbol{\phi}_i'\mathbf{f}_t$ capture latent time-varying heterogeneity through a factor model of rank $r \ll \min\{N, T\}$, where the common factors $\mathbf{f}_t$ may be arbitrarily correlated with both the treatment and the outcome process.
\end{remark}
\noindent
We assume that the covariates $\Xit$ are strictly exogenous in the sense of being conditionally mean independent of all lagged outcomes and treatments given the time-varying ommitted variables and lagged regressors. 
These restrictions ensure that the nuisance functions $l_0(\Xit)$ and $m_0(\Xit)$ can be estimated independently of the latent structure. 
The panel DML-IFE model provides a general framework that nests several widely used panel data models as special cases, depending on the specification of the number of factors $r$ and the structure of the latent components. Below we discuss several notable examples.

\paragraph{(i) Static DML with individual fixed effects (FE).}

When $r = 1$ and the common factor is constant across time, i.e., $f_t = 1$ for all $t$, the interactive fixed effects reduce to unit-specific constants:
\begin{equation}
\lambda_i^{\prime} f_t = \alpha_i, \qquad \Gamma_i^{\prime} f_t = \eta_i.
\end{equation}
In this case, the model simplifies to the static DML model with individual fixed effects as studied in \cite{clarke2025double}. The unobserved heterogeneity is entirely absorbed by unit-specific intercepts, and no temporal variation in the factor structure is permitted.

\paragraph{(ii) Two-way fixed effects (TWFE) under DML.}

When $r = 2$, we may recover the traditional two-way fixed effects model by specifying the latent factors and loadings as:
\begin{equation}
\vector{f}_t = \begin{bmatrix} 1 \\ \xi_t \end{bmatrix}, \quad 
\vector{\lambda}_i = \begin{bmatrix} \alpha_i \\ 1 \end{bmatrix}, \quad 
\boldsymbol{\Gamma}_i = \begin{bmatrix} \eta_i \\ 1 \end{bmatrix}.
\end{equation}
Under this parametrization, the interactive fixed effects reduce to:
\begin{equation}
\vector{\lambda}_i^{\prime} \vector{f}_t = \alpha_i + \xi_t, \qquad 
\boldsymbol{\Gamma}_i^{\prime} \vector{f}_t = \eta_i + \xi_t,
\end{equation}
which corresponds to the standard two-way fixed effects structure, where $\alpha_i$ and $\eta_i$ capture unit-specific heterogeneity and $\xi_t$ capture common time shocks in the outcome and treatment equations, respectively. However, the DML framework accommodates this structure while still allowing for high-dimensional covariate adjustments via machine learning.



Stacking Equations~\eqref{eq:basic_model1}-\eqref{eq:basic_model3}  over $t$ for each individual $i$, we obtain
\begin{align}
\boldsymbol{Y}_i &= \boldsymbol{V}_i\theta_0  + \boldsymbol{l}(\textbf{X}_i) + \mathbf{F} \lambda_i + \boldsymbol{U}_i, \label{eq:stacked_model_Y} \\
\boldsymbol{V}_i  &= \boldsymbol{D}_i - \boldsymbol{m}(\textbf{X}_i) - \mathbf{F} \vector{\phi}_i , \label{eq:stacked_model_D} \\
\mathbf{X}_i &=\mathbf{F} \Gamma_i^{\prime} + \mathbf{E}_i
\label{eq:stacked_model_X}
\end{align}
where $\boldsymbol{Y}_i = (Y_{i1}, \dots, Y_{iT})'$, 
$\boldsymbol{D}_i = (D_{i1}, \dots, D_{iT})'$, 
$\boldsymbol{U}_i = (U_{i1}, \dots, U_{iT})'$, 
$\boldsymbol{V}_i = (V_{i1}, \dots, V_{iT})'$, 
$\mathbf{F} = (f_1^0, \dots, f_T^0)'$, and
$\lambda_i, \Gamma_i \in \mathbb{R}^r$.
Here, $\mathbf{F}$ is a $T \times r$ matrix of latent common factors shared across individuals, while $\vector{\lambda}_i$ and $\vector{\phi}_i$ are $r \times 1$ vectors of unit-specific factor loadings in the outcome and treatment equations, respectively. $\vector{\phi}_i \in \mathbb{R}^{r}$ is the high-dimensional factor loadings. The vectors $\boldsymbol{U}_i$, $\boldsymbol{V}_i$  and matrix $\mathbf{E}_i$ collect idiosyncratic errors over $t$.  Finally, the functions $\boldsymbol{l}(\textbf{X}_i)$ and $\boldsymbol{m}(\textbf{X}_i)$ denote the $T \times 1$ vectors of pointwise evaluations of the nuisance functions at each time period, that is,
\[
\boldsymbol{l}(\textbf{X}_i) =
\begin{pmatrix}
l_0(\textbf{X}_{i1}) \\ l_0(\textbf{X}_{i2}) \\ \vdots \\ l_0(\textbf{X}_{iT})
\end{pmatrix}
\qquad
\boldsymbol{m}(\textbf{X}_i) =
\begin{pmatrix}
m_0(\textbf{X}_{i1}) \\ m_0(\textbf{X}_{i2}) \\ \vdots \\ m_0(\textbf{X}_{iT})
\end{pmatrix}.
\]

\subsection{Data Transformation}

For low-dimensional cases where $T>p\geq r$, \citet{pesaran2006estimation} shows that the key to estimating (\ref{eq:stacked_model_Y}) is through a projection matrix. Specifically, in Oracle scenarios where $\mathbf{F}$ is known, the projection matrix for any point in $\mathbb{R}^r$ onto the orthonormal complement of $\mathcal{L}_{\mathbf{F}}=\{\mathbf{F}\mathbf{v}: \mathbf{v}\in\mathbb{R}^r\}$, the $r$-dimensional column space of $\mathbf{F}$, is $\Pi = \mathbf{I}_T - \mathbf{F}(\mathbf{F}'\mathbf{F})^{-1}\mathbf{F}'$. It follows that $\Pi\boldsymbol{Y}_i = \Pi\boldsymbol{V}_i\theta_0  + \Pi \boldsymbol{l}_0(\textbf{X}_i)+\Pi\boldsymbol{U}_i$ with the interactive effects transformed out because, by definition, $\Pi\boldsymbol{F}=\mathbf{0}$. But because $\mathbf{F}$ and thus $\Pi$ are unavailable to the analyst, \citet{pesaran2006estimation} proposed instead the `proxy' projection matrix
\begin{equation}
 \widehat{\Pi} = \mathbf{I}_T - \overline{\mathbf{X}}(\overline{\mathbf{X}}'\overline{\mathbf{X}})^{-}\overline{\mathbf{X}}',
 \label{eq:proxy_proj_low}
\end{equation}
where $\overline{\mathbf{X}}=N^{-1}\sum_{i=1}^N\mathbf{X}_i$ contains the cross-sectional averages of all $p$ predictors at all $T$ waves, and $(\overline{\mathbf{X}}'\overline{\mathbf{X}})^{-}$ is the generalized inverse of $\overline{\mathbf{X}}'\overline{\mathbf{X}}$. Heuristically, it follows under (\ref{eq:stacked_model_X}) that $\overline{\mathbf{X}}\approx\boldsymbol{F} \Gamma^{\prime}$, where $\Gamma=\E[\Gamma_i]$, so that $\widehat{\Pi}\overline{\mathbf{X}}\approx\widehat{\Pi}\mathbf{F}\boldsymbol{\Gamma}'=\mathbf{0}_{T\times p}$ implies $\widehat{\Pi}\mathbf{F}=\mathbf{0}_{T\times r}$ under regularity conditions.
Applying the transformation to (\ref{eq:stacked_model_Y})-(\ref{eq:stacked_model_D}) gives working model
\begin{align}
\widehat{\Pi}\boldsymbol{Y}_i &= \widehat{\Pi}\boldsymbol{V}_i\theta_0  + \widehat{\Pi}\boldsymbol{l}(\textbf{X}_i)  + \widehat{\Pi}\boldsymbol{U}_i, \label{eq:transformed_model_Y} \\
\widehat{\Pi}\boldsymbol{V}_i  &= \widehat{\Pi}\boldsymbol{D}_i - \widehat{\Pi}\boldsymbol{m}(\textbf{X}_i)  .\label{eq:transformed_model_D}
\end{align}

\noindent For high-dimensional cases where $p>T$, the approach above breaks down because $\widehat{\Pi}=\boldsymbol{0}$. \citet{rucker2025} instead propose a novel approach based on the implication of (\ref{eq:stacked_model_X}) that the underlying dimension of the column space of $\overline{\boldsymbol{X}}$ is effectively $r < T$, even in asymptotic settings where $p=p(n)$ increases faster than $T$. Letting $\bm{\Phi}=(\boldsymbol{\varphi}_1,\ldots,\boldsymbol{\varphi}_r)$ be a matrix formed by the $r$ eigenvectors of $\boldsymbol{\Sigma}=\boldsymbol{\Gamma}\boldsymbol{\Gamma}'$, the column space of $\mathbf{W}=\overline{\mathbf{X}}\bm{\Phi}$ is shown using the singular value decomposition to be $\mathcal{L}_{\mathbf{F}}$.  Hence, the high-dimensional projection is $\Pi = \mathbf{I}_T - \mathbf{W}(\mathbf{W}'\mathbf{W})^{-1}\mathbf{W}'$.

In practice, unknown $r$ is estimated by first obtaining the eigenvalues $\widehat{\varphi}_1\geq\ldots\geq\widehat{\varphi}_p$ of $\widehat{\boldsymbol{\Sigma}}=T^{-1}\sum_{t=1}^T\bar{\mathbf{X}}_t\bar{\mathbf{X}}'_t$ ($\approx\boldsymbol{\Sigma}$) and then calculating
\begin{equation}
\widehat{r}=\sum_{i=1}^pI(\widehat{\varphi}_i\geq\tau),
\label{eq:factor_dim}
\end{equation}
where $\tau=\alpha\widehat{\varphi}_1$ and the user sets $\alpha=0.01,0.05$ or another small number.  In other words, eigenvectors with singular values close to zero, taken to be those explaining less than $100\cdot\alpha \%$ of the total variation, are excluded from the basis set. The proxy $\widehat{\Pi} $ is then 
\begin{equation}
\widehat{\Pi} = \mathbf{I}_T - \widehat{\mathbf{W}}(\widehat{\mathbf{W}}'\widehat{\mathbf{W}})^{-}\widehat{\mathbf{W}}',
\label{eq:proxy_proj_high}
\end{equation}
where $\widehat{\mathbf{W}}=\overline{\mathbf{X}}\widehat{\bm{\Phi}}$, $\widehat{\bm{\Phi}}=(\widehat{\boldsymbol{\varphi}}_1,\ldots,\widehat{\boldsymbol{\varphi}}_{\widehat{r}})$ and $\widehat{\boldsymbol{\varphi}}_j$ is the eigenvector corresponding to eigenvalue $\widehat{\varphi}_j$ for $j=1,\ldots,\widehat{r}$ above.
\subsection{Assumptions}
The assumptions needed to identify the panel DML-IFE model are as follows:

\begin{assumption}\label{item:asm_process} 
\rm{
    \textbf{(Panel process)} Let $\xi_{it}$ represent the entire set of omitted confounders for individual $i$ at time $t$. Then $\{\Yit,\Dit,\Xit,\xi_{it}: t\in\mathbb{N}\}$ satisfies the conditions below.
    \begin{enumerate}[label=(\alph*)]
    \item $\Xit,\xi_{it} \perp\!\!\!\perp \{Y_{is},D_{is} : s<t\} \mid \{\mathbf{X}_{is},\xi_{is} : s<t\}$, \label{item:asm_process1}
    \item $Y_{it}, D_{it}  \perp\!\!\!\perp \{Y_{is}, D_{is}, \textbf{X}_{is},\xi_{is} : s < t \} \mid  {X}_{it},\xi_{it}$, \label{item:asm_process2}
    \end{enumerate}
    where $\perp\!\!\!\perp$ indicates conditional (stochastic) independence.
}
\end{assumption}

\begin{remark}
The first assumption is that all of the confounding variables, observed or omitted, are {\bf strictly exogenous} in that the outcome and treatment lags are not predictive of the confounding variables (observed and omitted) given confounding history; this ensures that, for the purposes of inference, all confounding can be treated as determined prior to \mbox{$t=1$}. The second assumption is that the panel model is {\bf static} in that the outcome and treatment are conditionally lag-independent given both sets of confounding variables at wave $t$. Failing to account for the omitted confounding leads to bias because $\mathbb{E}[\Uit\mid\Dit,\Xit]\neq 0$ and $\mathbb{E}[\Vit\mid\Xit]\neq 0$.
\end{remark}

\begin{assumption}\label{item:asm_separability} 
\rm{
    \textbf{(Additive separability)} 
    \begin{enumerate}[label=(\alph*)]
    \item $\Yit =\theta_0\Vit +\widetilde{l}_0\left(\Xit,\xi_{it}\right)+\Uit$, where $\mathbb{E}[\Uit\mid\Dit, \Xit,\xi_{it}]=0$, $\widetilde{l}_0(\Xit,\xi_{it})=l_0(\Xit)+\alpha_{it}$ and fixed effect $\alpha_{it}=\alpha(\xi_{it})$, \label{item:asm_separability1}
    \item $\Vit =\Dit -\widetilde{m}_0\left(\Xit,\xi_{it}\right)$, where $\mathbb{E}[\Vit\mid\Xit,\xi_{it}]=0$, $\widetilde{m}_0(\Xit,\xi_{it})=m_0(\Xit)+\gamma_{it}$ and fixed effect $\gamma_{it}=\gamma(\xi_{it})$.\label{item:asm_separability2}
    \end{enumerate}
}
\end{assumption}

\begin{remark}
The separation of the treatment effect in the outcome model from the confounding effects follows if there is no treatment effect heterogeneity with respect to $\Xit,\xi_{it}$ (c.f.\ \citet[sec.~3.2]{clarke2025double}). The separability of the observed and omitted confounding effects due to \citet{rucker2025} has already been discussed. 
\end{remark}

\begin{assumption}\label{item:asm_latent} 
\rm{
    \textbf{(Latent factor model for fixed effects)} 
    \begin{enumerate}[label=(\alph*)]
    \item $\Xit=\boldsymbol{\eta}(\xi_{it})+\mathbf{E}_{it}$, where $\boldsymbol{\eta}(\xi_{it})=\Gamma_i^{\prime}\mathbf{f}_t$ and $\mathbb{E}[\mathbf{E}_{it} \mid \Xit, \Gamma_i,\vector{f}_t]  = {\bf 0}$, \label{item:asm_latent1}
    \item  $\Yit =\theta_0\Vit +l_0\left(\Xit\right)+\alpha(\xi_{it})+\Uit$, where $\alpha(\xi_{it})=\boldsymbol{\lambda}_i^{\prime}\vector{f}_t$ and $\mathbb{E}[U_{it} \mid \Dit, \Xit, \boldsymbol{\lambda}_i,\vector{f}_t]  = 0$,\label{item:asm_latent2}
\item $V_{i t}=\Dit - m_0\left(\Xit\right) - \gamma(\xi_{it})$, where
    $\gamma\left(\xi_{it}\right)=\boldsymbol{\phi}_i^{\prime}\mathbf{f}_t$ and
    $\mathbb{E}[V_{it} \mid \Xit, \boldsymbol{\phi}_{i},\vector{f}_t]  = 0$.\label{item:asm_latent3}
    \end{enumerate}
}
\end{assumption}
\begin{remark}
The latent factors $\vector{f}_t$ vary over time but the time pattern is common across individuals and processes; the loadings indexing the influence of these factors, however, vary between individuals and differ for the outcome, treatment and observed confounding variables. Assumption~\ref{item:asm_latent}.\ref{item:asm_latent1}  plays a key role in identifying the time-varying fixed effects.
\end{remark}

\begin{assumption}\label{item:asm_factors} 
\rm{
    \textbf{(Identification of factors)} 
    \begin{enumerate}[label=(\alph*)]
        \item $\mathbf{F}^{\prime}\mathbf{F}/T = \mathbf{I}_{r\times r}$,\label{item:asm_factors1} 
        \item $E\left\|\vector{f}_t\right\|^4 \leq C \text { and }  T^{-1}\mathbf{F}^{\prime}\mathbf{F} \xrightarrow{p} \boldsymbol{\Sigma}^0_F$ as $T \rightarrow \infty$ for some non-random positive definite matrix $\boldsymbol{\Sigma}_F^0$.\label{item:asm_factors2} 
    \end{enumerate}
}
\end{assumption}

\begin{remark}
  Part~(a) selects the unique normalisation under which $\mathbf{F}$ has 
  orthonormal columns, such that $\boldsymbol{\Sigma}_F^0 = \mathbf{I}_r$, needed to connect the column spaces of $\mathbf{F}$, $\bar{\mathbf{X}}$ and $\mathbf{W}$. 
  The substantive content of Part~(b) is the fourth-moment bound $\mathbb{E}\|\vector{f}_t\|^4\leq C$. The bound holds for bounded factors (e.g.\ 
  cohort or period dummies), sub-Gaussian factors, and any factor process 
  satisfying the strong mixing conditions of Assumption~\ref{item:asm_process}.
\end{remark}

\begin{assumption}\label{item:asm_loadings} 
\rm{
    \textbf{(Identification of factor loadings)} 
    \begin{enumerate}[label=(\alph*)]
        \item  $E\left\|\vector{\lambda}_i\right\|^4 \leq C \text { and }  N^{-1}\sum_{i=1}^N \vector{\lambda}_i\vector{\lambda}_i^{\prime} \xrightarrow{p} \boldsymbol{\Sigma}_\lambda$ as $N \rightarrow \infty$ for some non-random positive definite matrix $\boldsymbol{\Sigma}_\lambda$.\label{item:asm_loadings1} 
        \item $E\left\|\boldsymbol{\Gamma}_i\right\|^4 \leq C \text { and }  N^{-1} \sum_{i=1}^N \boldsymbol{\Gamma}_i\boldsymbol{\Gamma}_i^{\prime} \xrightarrow{p} \boldsymbol{\Sigma}_\Gamma$ as $N \rightarrow \infty$ for some non-random positive definite matrix $\boldsymbol{\Sigma}_\Gamma$.\label{item:asm_loadings2} 
        \item $E\left\|\vector{\phi}_i\right\|^4 \leq C \text { and }  N^{-1} \sum_{i=1}^N \vector{\phi}_i\vector{\phi}_i^{\prime} \xrightarrow{p} \boldsymbol{\Sigma}_\phi$ as $N \rightarrow \infty$ for some non-random positive definite matrix $\boldsymbol{\Sigma}_\phi$.\label{item:asm_loadings3} 
       \item Following the normalisation in factors, we impose the following assumption
on the mean loading matrix $\bar{\boldsymbol{\Gamma}}=\mathbb{E}\left[\boldsymbol{\Gamma}_i\right] \in \mathbb{R}^{p \times r}$.
Let $\varphi_1(\mathbf{A}) \ge \dots \ge \varphi_r(\mathbf{A})$ denote the ordered eigenvalues
of an $r \times r$ matrix $\mathbf{A}$. We assume the eigenvalues of
$\mathbf{A} = \bar{\boldsymbol{\Gamma}}^{\top} \bar{\boldsymbol{\Gamma}} / p$ are bounded and strictly bounded
away from zero: there exist constants $0 < c_{\min} \le c_{\max} < \infty$ such that
$$\varphi_r(\bar{\boldsymbol{\Gamma}}^{\top} \bar{\boldsymbol{\Gamma}} / p) \ge c_{\min}
  \quad \text{and} \quad
  \varphi_1(\bar{\boldsymbol{\Gamma}}^{\top} \bar{\boldsymbol{\Gamma}} / p) \le c_{\max}.$$
Equivalently $\sqrt{c_{\min}\,p}\le\sigma_r(\bar{\boldsymbol{\Gamma}})\le\sigma_1(\bar{\boldsymbol{\Gamma}})\le\sqrt{c_{\max}\,p}$,
so the factors are pervasive: $\sigma_r(\bar{\boldsymbol{\Gamma}})\asymp\sqrt{p}$.\label{item:asm_loadings4}
    \end{enumerate}
}
\end{assumption}

\begin{remark}
  Parts~(a)--(c) are standard moment and law-of-large-numbers (LLN) conditions 
  analogous to Assumption~C of \citet{bai2009panel}: the positive 
  definiteness of $\boldsymbol{\Sigma}_\lambda$, $\boldsymbol{\Sigma}_\Gamma$, 
  $\boldsymbol{\Sigma}_\phi$ rules out degenerate factor structures in which 
  some factor direction is systematically unloaded, and the fourth-moment 
  bounds ensure concentration of sample covariances at the rate required 
  in the proof of Lemma. Part~(d) is the key condition enabling the CCE proxy estimator. Since 
  $\bar{\boldsymbol{X}}_t\approx\boldsymbol{\Gamma}^\top\vector{f}_t$ for large $N$, the 
  lower bound $\varphi_r(\boldsymbol{\gamma}^\top\boldsymbol{\gamma}/p)\geq c_{\min}$ 
  ensures all $r$ factors are recoverable from $\bar{\mathbf{X}}$, whilst 
  the upper bound prevents ill-conditioning when $p$ is large. 
\end{remark}
\begin{assumption}[\bf Factor loading identification]\label{asm:loadings_ident}
  \rm
  Under the pervasiveness normalisation of
  Assumption~\ref{item:asm_loadings}\ref{item:asm_loadings4}, the cross-sectional
  mean loading $\bar{\boldsymbol{\Gamma}} = N^{-1}\sum_{i=1}^N\boldsymbol{\Gamma}_i \in \mathbb{R}^{p\times r}$
  satisfies, with probability approaching one,
  $$\sigma_r(\bar{\boldsymbol{\Gamma}}) \ge c_\Gamma\sqrt{p}, \qquad c_\Gamma := \sqrt{c_{\min}} > 0.$$
  This follows from Assumption~\ref{item:asm_loadings}\ref{item:asm_loadings4} applied
  to the sample mean $\bar{\boldsymbol{\Gamma}}$ (which converges to $\mathbb E[\boldsymbol{\Gamma}_i]$ by the LLN
  under the fourth-moment bound), and is what makes the projection $\widehat\Pi$
  well-defined with $\mathcal{R}(\overline{X}) \supseteq \mathcal{R}(F)$ asymptotically.
\end{assumption}

\begin{assumption}[\bf Machine learning rate conditions]\label{asm:ml_rate}
  \rm
  Let $\widehat{l}_k$ and $\widehat{m}_k$ denote the predictions of the true population parameters $l_0$ and $m_0$
  trained on the complement of fold $k$.
  There exists a sequence $\delta_{NT}\to 0$ such that, for each $k=1,\ldots,K$,
  \begin{equation}\label{eq:ml_rate}
    \bigl\|\widehat{l}_k - l_0\bigr\|_{L_2}
    = O_p(\delta_{NT}), \qquad
    \bigl\|\widehat{m}_k - m_0\bigr\|_{L_2}
    = O_p(\delta_{NT}),
  \end{equation}
  where $\|\cdot\|_{L_2}$ denotes the $L_2(P)$-norm.
  Moreover, $\delta_{NT} = o\bigl((NT)^{-1/4}\bigr)$, so that
  $\delta_{NT}^2 = o\bigl((NT)^{-1/2}\bigr)$.
\end{assumption}

\begin{remark}[Achievability]
  The rate $o\bigl((NT)^{-1/4}\bigr)$ is equivalent to the $o(N^{-1/4})$ rate used
  in \citet[][Assumption~3.2]{chernozhukov2018double} for
  effective sample size $NT$.  It is achieved, for example, by Lasso under
  sparsity (see the $\ell_1$ rates in \citet{rucker2025},
  Theorem~6.1), by boosting, and by neural networks under smoothness conditions
  \citep{farrell2021deep}. Achieving this rate for a given application characterises the wide class of machine learning algorithms which can be used for this procedure.
\end{remark}

\begin{assumption}[\bf Dimension growth conditions]\label{asm:dimension}
  \rm
  The dimensions $(N,T,p)$ satisfy, as $(N,T)\to\infty$:
  \begin{enumerate}[label=(\alph*)]
    \item $r = O(1)$ (the number of latent factors is small and fixed);\label{asm:dim_r}
    \item $(T+p)^2\log^2(T+p) = o(NT)$;\label{asm:dim_proj}
          \emph{This is the condition that emerges from demanding
          the  projection-error bias $O_p\bigl((T+p)\log(T+p)/N\bigr)$
          is $o_p\bigl((NT)^{-1/2}\bigr)$.
          For $p = O(T)$ it reduces to $T\log^2 T = o(N)$.)}
    \item $\delta_{NT}^2\sqrt{NT}\to 0$, ensuring the product of ML errors is
          negligible at the $\sqrt{NT}$ rate.\label{asm:dim_ml}
    \item the spectral gap condition
  $\sigma_r(\mathbf{F}\bar{\boldsymbol{\Gamma}}^\top) - \sigma_{r+1}(\overline{\mathbf X})
  \geq c_\Gamma\sqrt T/2$ w.p.a.1 follows from
  $\sigma_r(\mathbf \mathbf{F}{\bar{\boldsymbol{\Gamma}}}^\top) = \sqrt T\sigma_r(\bar{\boldsymbol{\Gamma}})\geq c_\Gamma\sqrt T$
  (Assumption~\ref{asm:loadings_ident}, normalisation
  $\mathbf F'\mathbf F = T\mathbf I_r$) and
  $\sigma_{r+1}(\overline{\mathbf X}) \leq \|\overline{\mathbf E}\|_{\mathrm{op}}
  = O_p(\sqrt{(T+p)\log(T+p)/N}) = o_p(\sqrt T)$ by
  Assumption~\ref{asm:dim_proj}.
  \end{enumerate}
\end{assumption}

\begin{assumption}[\bf Concentration of idiosyncratic covariate errors]\label{asm:E_conc}
  \rm
  The idiosyncratic covariate errors $\mathbf E_i = \mathbf X_i - \mathbf F\Gamma_i'
  \in\mathbb R^{T\times p}$ are independent across $i$ with
  $\mathbb E[\mathbf E_i]=\mathbf 0_{T\times p}$; its $(t,j)$ entry is denoted $E_{i,tj}$ and equals the
  $j$-th coordinate of the time-$t$ error vector $\mathbf E_{it}$
  (so the $t$-th row of $\mathbf E_i$ is $\mathbf E_{it}'$ and the $j$-th column
  collects the $j$-th covariate's error over $t=1,\ldots,T$).
  The errors satisfy the following conditions, in which
  $\sigma_E^2$ and $M_E$ are constants that do \emph{not} depend on $(N,T,p)$.
  \begin{enumerate}[label=(\alph*)]
    \item \emph{(Bounded second moments and weak dependence.)}
      $\mathbb E[E_{i,tj}^2]\leq\sigma_E^2<\infty$ for all $t,j$, and the rows
      and columns of $\mathbf E_i$ are weakly dependent in the sense that the
      per-unit second-moment matrices have operator norms growing at most
      linearly in the corresponding dimension:
      \begin{equation}\label{eq:E_var_proxy}
        \bigl\|\mathbb E[\mathbf E_i \mathbf E_i']\bigr\|_{\mathrm{op}} \leq \sigma_E^2\,p,
        \qquad
        \bigl\|\mathbb E[\mathbf E_i' \mathbf E_i]\bigr\|_{\mathrm{op}} \leq \sigma_E^2\,T .
      \end{equation}
    \item \emph{(Tail condition.)} Either $\max_i\|\mathbf E_i\|_{\mathrm{op}}\leq M_E$
      almost surely, or the rows of $\mathbf E_i$ are sub-Gaussian with parameter
      $\sigma_E$, so that the Matrix Bernstein inequality applies to
      $\overline{\mathbf E} = N^{-1}\sum_i \mathbf E_i$.
  \end{enumerate}
  \begin{remark}\normalfont
  The bounds in~\eqref{eq:E_var_proxy} are the correct objects for the matrix
  Bernstein inequality: because $[\mathbf E_i\mathbf E_i']_{ts}=\sum_{j=1}^p E_{i,tj}E_{i,sj}$
  aggregates over the $p$ covariates, $\|\mathbb E[\mathbf E_i\mathbf E_i']\|_{\mathrm{op}}$ is of
  order $p$ (and symmetrically $\|\mathbb E[\mathbf E_i'\mathbf E_i]\|_{\mathrm{op}}$ of order $T$)
  whenever the per-element variances are $O(1)$ and cross-correlations are
  summable; it is the constant $\sigma_E^2$, not the matrix norm itself, that is
  bounded uniformly in $(N,T,p)$.  Under independence across $i$, the
  variance proxy of the average $\overline{\mathbf E}$ is then
  $N^{-1}\sigma_E^2\max(T,p)\asymp\sigma_E^2(T+p)/N$.
\end{remark}
\end{assumption}

\begin{assumption}[\bf Moment and weak dependence conditions]\label{asm:moments}
  \rm
  \begin{enumerate}[label=(\alph*)]
  \item $\E\|\Uit\|^{4+\delta} \leq C$ and $\E\|\Vit\|^{4+\delta}\leq C$
          for some $\delta>0$ and a universal constant $C<\infty$.\label{asm:moment4}
 \item For each unit $i$, the joint sequence $\{(U_{it},V_{it})\}_{t=1}^T$ is
          fourth-order stationary, and the score-contribution process
          $\{\widetilde V_{it}\widetilde U_{it}\}_t$ has absolutely summable
          long-run covariances:
          $\sum_{k=-\infty}^{\infty}
           \bigl|\mathrm{Cov}(\widetilde V_{it}\widetilde U_{it},\,
                 \widetilde V_{i,t+k}\widetilde U_{i,t+k})\bigr| < \infty.$
          A sufficient primitive condition is that $(U_{it},V_{it})$ is jointly
          $\alpha$-mixing with $\sum_k\alpha(k)^{\delta/(4+\delta)}<\infty$,
          combined with~\ref{asm:moment4}.\label{asm:mixing}
    \item Units are cross-sectionally independent: $(W_{it})_{t\geq1}$ and
          $(W_{jt})_{t\geq1}$ are independent for $i\neq j$.\label{asm:ci}
   \item (Relevance / no weak identification.)
          The projected treatment residual has variation bounded away from
          zero and above:
          $0 < c_V \leq \E[\widetilde V_{it}^2] \leq C_V < \infty$,
          where $\widetilde V_{it} = [\Pi\boldsymbol V_i]_t$.
          Equivalently, after projecting out the latent factors, the treatment
          retains non-degenerate idiosyncratic variation.\label{asm:relevance}
  \end{enumerate}
\end{assumption}
\begin{remark}
  Part~(a) defines the $4{+}\delta$ moment on $\Uit$ and $\Vit$. It
  implies a finite $4{+}\delta$ moment on the score product 
  $\widetilde{V}_{it}\widetilde{U}_{it}$, which is sufficient to verify 
  that the Lyapunov ratio tends to zero as $N\to\infty$.

   Part~(b) allows serial correlation within each unit. Absolute summability of
  the long-run covariances of $\{\widetilde V_{it}\widetilde U_{it}\}$ is
  equivalent to its spectral density being bounded and continuous at frequency
  zero, and ensures the long-run variance
  $\mathcal{V}=\sum_{k=-\infty}^{\infty}\mathrm{Cov}(\widetilde{V}_{it}
  \widetilde{U}_{it},\widetilde{V}_{i,t+k}\widetilde{U}_{i,t+k})$
  is well-defined and finite. When the score product is 
  serially uncorrelated, $\mathcal{V}$ reduces to 
  $\mathbb{E}[\widetilde{V}_{it}^2\widetilde{U}_{it}^2]$, in which case a 
  simple heteroskedasticity-robust variance estimator suffices; otherwise a 
  HAC estimator is required for 
  $\widehat{\sigma}^2$.

  Part~(c) rules out cross-sectional dependence, enabling the 
  Lindeberg--Feller CLT to be applied across units $i=1,\ldots,N$ 
  conditionally on $\mathbf{F}$. This is the standard assumption in the 
  interactive fixed-effects literature \citep{bai2009panel,moon2017dynamic}. 
  Relaxing Part~(c) to weak cross-sectional dependence is feasible in principle but would require a 
  mixing-based CLT and is left for future work.
\end{remark}
\section{Panel DML-IFE Estimation Procedure}\label{sec:algorithm}

Suppose we have obtained first-step estimators $\widehat{\vector{l}}(\cdot)$ and $\widehat{\vector{m}}(\cdot)$
of nuisance functions $\vector{l}_0(\cdot)$ and $\vector{m}_0(\cdot)$ using flexible nonparametric or semiparametric estimation or machine learning.
A plug-in estimator for the treatment effect $\theta_0$ can be constructed as
the solution to the following score:
\begin{equation}\label{eq:foc}
  \frac{1}{NT}\sum_{i=1}^{N}\sum_{t=1}^{T}
  \widehat{\psi}_{it}\!\left(\theta;\widehat{\vector{l}},\widehat{\vector{m}}\right) = 0,
\end{equation}
where (see Appendix A.1)
\[
  \widehat{\psi}_{it}(\theta;\hat{\vector{l}},\hat{\vector{m}})
  = [\widehat{\Pi}\{ \mathbf{D}_i-\widehat{\vector{m}}(\mathbf{X}_{i})\}]_t
            [\widehat{\Pi}\{\mathbf{Y}_i - (\mathbf{D}_i-\widehat{\vector{m}}(\mathbf{X}_i))\theta - \widehat{\vector{l}}(\mathbf{X}_{i})\}]_t /\widehat{\sigma}^2_{\tilde{u}},
\]
$[\boldsymbol{A}]_t$ is used to denote co-ordinate $t$ of arbitrary vector $\boldsymbol{A}$, and $\widehat{\sigma}^2_{\tilde{u}}$ is an estimate of the variance of residual $\tilde{U}_{it}=[\widehat{\Pi}\boldsymbol{U}_i]_t$ from (\ref{eq:transformed_model_Y}).

The oracle version of this score
\begin{equation}\label{eq:score}
  \psi_{it}(\theta_0;\vector{l}_0,\vector{m}_0)
  = [\Pi_0( \boldsymbol{D}_i-\vector{m}_0(\mathbf{X}_{i}))]_t
            [\Pi_0(\boldsymbol{Y}_{it} - (\boldsymbol{D}_{it}-\vector{m}_0(\mathbf{X}_i))\theta_0 - \vector{l}_0(\mathbf{X}_i))]_t /\sigma^2_{\tilde{u}}
\end{equation}
is locally efficient in the sense of being fully efficient if residuals $\tilde{U}_{it}$ are uncorrelated and homoscedastic. At the true values $(\vector{l}_0,\vector{m}_0,\sigma^2_{\tilde{u}})$, the factor structure of the omitted variables 
drops out by construction because $\Pi\mathbf{F}=\mathbf{0}$,
so the first factor reduces to
$[\Pi_0\boldsymbol{V}_i]_t=\widetilde{V}_{it}$ and the inner residual of the second factor
satisfies $[\Pi_0(\boldsymbol{Y}_i- (\boldsymbol{D}_{it}-\vector{m}_0(\mathbf{X}_i))\theta_0-l_0(\mathbf{X}_i))]_t
= [\Pi_0 \boldsymbol{U}_i]_t=\widetilde{U}_{it}$.

Inference on $\theta_0$ follows from a semiparametric Taylor expansion of
\eqref{eq:foc} around the true values $(\theta_0,l_0,m_0)$. Following the presentation by \citet[p.~14]{ahrens2026}, this can be written
\begin{align}
\sqrt{NT}(\widehat\theta-\theta_0)
  &= -\!\left[\frac{1}{NT}\!\sum_{i,t}
       \frac{\partial\psi_{it}}{\partial\theta}\bigg|_0
     \right]^{\!-1}
     \underbrace{%
       \frac{1}{\sqrt{NT}}\sum_{i,t}\widetilde{V}_{it}\widetilde{U}_{it}
     }_{\text{CLT}} \notag\\
  &\quad +\left[\frac{1}{NT}\!\sum_{i,t}
       \frac{\partial\psi_{it}}{\partial\theta}\bigg|_0
     \right]^{\!-1}
     \underbrace{%
       \frac{1}{NT}\!\sum_{i,t}\!\left[
         \frac{\partial\psi_{it}}{\partial l}\bigg|_0\!(\widehat{\vector{l}}-\vector{l}_0)
         +\frac{\partial\psi_{it}}{\partial m}\bigg|_0\!(\widehat{\vector{m}}-\vector{m}_0)
       \right]
     }_{(\star)\colon\;\text{first-order impact of nuisance estimation}}
     \notag\\
  &\quad +\;\sqrt{NT}\times(\text{higher order terms}).
  \label{eq:taylor_ife}
\end{align}
Note that $\widehat{\Pi}$ can be replaced by $\Pi_0$ because $\sqrt{NT}(\widehat{\Pi}-\Pi_0)=o_p(1)$ under Assumption~\ref{asm:dimension}(b), with $\widehat{\sigma}^2_{\tilde{u}}$ is similarly replaced by $\sigma^2_{\tilde{u}}$ (and then suppressed) because it is trivially $\sqrt{NT}$ consistent.

Because $\psi_{it}$ is linear in $\theta$, the Taylor expansion is exact in the
$\theta$ direction: no $(\widehat\theta-\theta_0)^2$ remainder appears.
The denominator factor equals
\begin{equation*}
  \frac{1}{NT}\sum_{i,t}\frac{\partial\psi_{it}}{\partial\theta}\bigg|_0
  = -\frac{1}{NT}\sum_{i,t}\widetilde{V}_{it}\widetilde{D}_{it},
  \qquad \widetilde{D}_{it}=[\Pi_0 \vector{D}_i]_t,
\end{equation*}
which converges in probability to
$J_0 = -\mathbb{E}[\widetilde{V}_{it}\widetilde{D}_{it}]
= -\mathbb{E}[\widetilde{V}_{it}^{\,2}]$
by the law of large numbers, the second equality following from
$\mathbb{E}[\widetilde V_{it}\cdot[\Pi_0 \vector{m}_0(\mathbf{X}_i)]_t]=0$.

The term $(\star)$ below captures the first-order impact of estimating the nuisance
functions $l_0$ and $m_0$. Expanding the Gateaux derivatives:
\begin{equation}\label{eq:star}
  (\star) =
  \underbrace{-\frac{1}{NT}\sum_{i,t}
    \widetilde{V}_{it}\,[\Pi_0(\widehat{\vector{l}}-\vector{l}_0)(\mathbf{X}_i)]_t}_{(\star_l)}
  \underbrace{-\frac{1}{NT}\sum_{i,t}
    [\Pi_0(\widehat{\vector{m}}-\vector{m}_0)(\mathbf{X}_i)]_t\,\widetilde{U}_{it}}_{(\star_m)},
\end{equation}
where $(\widehat{\vector{l}}-\vector{l}_0)(\mathbf{X}_i)=\widehat{\vector{l}}(\mathbf{X}_i)-\vector{l}_0(\mathbf{X}_i)$ and $(\widehat{\vector{m}}-\vector{m}_0)(\mathbf{X}_i)=\widehat{\vector{m}}(\mathbf{X}_i)-\vector{m}_0(\mathbf{X}_i)$. In general, $(\star)$ diverges due to two sources of bias.
Regularisation bias arises because flexible estimators of $l_0$ and
$m_0$ are biased in finite samples, so neither summand in \eqref{eq:star} is
mean zero in general.
Overfitting bias arises because, if the same sample is used to form
$\widehat{\vector{l}},\widehat{\vector{m}}$ and to evaluate the score, the estimation errors
$\widehat{\vector{l}}-\vector{l}_0$ and $\widehat{\vector{m}}-\vector{m}_0$ become correlated with the projected
residuals $\widetilde{V}_{it}$ and $\widetilde{U}_{it}$.

The DML-IFE score eliminates $(\star)$ through two devices.
First, Neyman orthogonality of the projected score eliminates $(\star)$ at the
population level. For any direction $\vector{h}=\widehat{\vector{l}}-\vector{l}_0$,
\begin{equation}\label{eq:neyman}
  \mathbb{E}\!\left[\widetilde{V}_{it}\,[\Pi_0 \vector{h}(\mathbf{X}_i)]_t\right]
  = \mathbb{E}\!\left\{
      [\Pi_0 \vector{h}(\mathbf{X}_i)]_t\,
      \underbrace{\mathbb{E}\!\left[\widetilde V_{it}\mid\mathbf{X}_i,\mathbf{F}\right]}_{=\;0}
    \right\} = 0,
\end{equation}
where the conditional expectation vanishes by strict exogeneity:
$\mathbb{E}[\widetilde{V}_{it}\mid\mathbf{X}_i,\mathbf{F}]
=[\Pi_0\,\mathbb{E}(\mathbf{V}_i\mid\mathbf{X}_i,\mathbf{F})]_t = 0$.
An identical argument applies to $(\star_m)$ via
$\mathbb{E}[\widetilde U_{it}\mid\mathbf{X}_i,D_{it},\mathbf{F}]=0$.
Second, cross-fitting controls overfitting bias by ensuring $\widehat{\vector{l}}_k$ and
$\widehat{\vector{m}}_k$, trained on the complement of fold $k$, are independent of the
evaluation-fold residuals.
Together, these two devices reduce $(\star)$ to $o_p((NT)^{-1/2})$, so \eqref{eq:taylor_ife} collapses to
\begin{equation}\label{eq:asymp}
  \sqrt{NT}(\widehat\theta-\theta_0)
  = -J_0^{-1}\cdot
    \frac{1}{\sqrt{NT}}\sum_{i,t}\widetilde{V}_{it}\widetilde{U}_{it}
    + o_p(1).
\end{equation}
The procedure for estimating the structural parameter $\theta_0$ and conducting statistical inference using the panel DML-IFE estimator is outlined below and summarised in Algorithm~\ref{alg:xtdml}. Assume IFE with no additive individual or time FE. The panel DML-IFE procedure follows the steps below: \\

\noindent \textbf{Step 1: Construct the projection matrix $\widehat{\Pi}$.} Calculate the average of the covariates over individuals, $\mathbf{X}=N^{-1}\sum_{i=1}^N\mathbf{X}_i$, so that we have $\widehat{\Pi} = \mathbf{I}_T - \widehat{\mathbf{W}}(\widehat{\mathbf{W}}'\widehat{\mathbf{W}})^{-}\widehat{\mathbf{W}}'$, where $\widehat{\mathbf{W}}=\overline{\mathbf{X}}\widehat{\boldsymbol{\Phi}}$, $\widehat{\boldsymbol{\Phi}}=(\widehat{\boldsymbol{\varphi}}_1,\ldots,\widehat{\boldsymbol{\varphi}}_{\widehat{r}})$ and $\widehat{\boldsymbol{\varphi}}_j$ is the eigenvector corresponding to eigenvalue $\widehat{\varphi}_j$ for $j=1,\ldots,\widehat{r}$ above.\\

\noindent \textbf{Step 2: Data transformation using the projection matrix.}
Transform the outcome, treatment and confounding variables by applying the projection matrix $\widehat{\Pi}$ on $\widetilde{Y} = \widehat{\Pi}Y$, $\widetilde{D} = \widehat{\Pi} D$, $\widetilde{{l}} = \widehat{\Pi} {l}(X)$, $\widetilde{{m}} = \widehat{\Pi}{m}(X)$.%
\footnote{Because we assume additive separability, so $\widehat{\Pi}\cdot [l(\boldsymbol{x}) + \Gamma_i'f_t]= \Pi_0 \cdot {l}(\boldsymbol{x}) + 0 = \tilde{l}(\boldsymbol{x}).$}\\

\noindent \textbf{Step 3: Panel DML-IFE estimator.} The estimation and inference about the structural parameter proceeds in the following steps:
\begin{enumerate}[label=\textbf{3.\arabic*}]
    \item \textbf{Block-k sample splitting and cross-fitting.} 
    Randomly partition the cross-sectional units $N$ into $K$ folds of the same size. For each fold \mbox{$k=1,\ldots,K$} denote $\Wk\subset\mathcal{W}$ as the estimation sample, and $\Wk^c$ as its complement, where $N_k\equiv |\Wk| = N/K$, $|\Wk^c|=N-N_k$. The folds are mutually exclusive and exhaustive such that $\Wk\cap\mathcal{W}_j=\Wk\cap\Wk^c=\varnothing$ and $\Wk\cup\Wk^c=\mathcal{W}_1\cup\ldots\cup\mathcal{W}_K=\mathcal{W}$.
    \item \textbf{Iteration.} For each fold:
    \begin{enumerate}
        \item \textbf{Learn the nuisance parameters}, $\bmeta = (l, m, g)$, using ML algorithms on complementary sample. A good practice is to use optimal (already tuned) hyperparameters.
        \item \textbf{Orthogonalisation against observed controls.} Use the predicted nuisances from previous step to  obtain the orthogonalised residuals, $\widetilde{U} = \widetilde{Y} - \widehat{l}(\widetilde{x})$ and $\widetilde{V}= \widetilde{D} - \widehat{m}(\widetilde{x})$ and construct a Neyman orthogonal score . 
        \item \textbf{Solution to moment condition.} Solve the empirical moment condition of the orthogonal score function using data in the main sample and obtain intermediate estimates of the structural parameter.
    \end{enumerate}
    \item \textbf{Final estimates.} Aggregate/average intermediate estimates based on the DML procedure.
\end{enumerate}

\noindent Steps~1 and~2 are new relative to standard DML procedures for panel data: they require obtaining a consistent estimate of the projection matrix, in the spirit of \citet{pesaran2006estimation}'s PCCE estimator, on the full sample to transform the available data. Step~3 is a modified version of \citet{clarke2025double}, where the transformed variables via the projection matrix handle the presence of the common factors, and are used to obtain the ML predictions of the nuisance functions. Finally, as standard practice in DML, the orthogonalised residuals, constructed from these ML predictions, enter the Neyman-orthogonal score function, from which the structural effect is recovered.


\begin{algorithm}[th!]
\SetAlgoLined
\SetKwData{Left}{left}
\SetKwData{This}{this}
\SetKwData{Up}{up}
\SetKwFunction{Union}{Union}
\SetKwFunction{FindCompress}{FindCompress}
\SetKwInOut{Input}{Input}
\SetKwInOut{Output}{Output}
\SetKwInOut{Require}{Require}
\SetKwInOut{State}{State}
\SetKwInOut{Initialize}{Initialize}
\caption{Panel DML Algorithm with Interactive Fixed Effects}\label{alg:xtdml}
\vspace{2mm}
\Require{Dataset with $N$ cross-sectional units, and $T\ge 2$ time periods. }
\Input{Data $\{Y_{it},D_{it},\vector X_{it}\}$; panel and time identifiers, cluster identifiers (optional).}
\Output{$\widehat{\theta}$, $\widehat{\Sigma}$, model RMSE, RMSE~of~$l$,~$m$,~$g$.}


\Initialize{Set number of folds $k\ge2$. Select the type of score function, and the double machine learning procedure. Assign base learners to each nuisance parameters $\bm{\eta} = \{l, m, g\}$ and define tuning settings.}

\vspace{2mm}

Construct the projection matrix $\widehat{\Pi} = \mathbf{I}_T - \widehat{\mathbf{W}}(\widehat{\mathbf{W}}'\widehat{\mathbf{W}})^{-}\widehat{\mathbf{W}}'$, where $\widehat{\mathbf{W}}=\overline{\mathbf{X}}\widehat{\boldsymbol{\Phi}}$, $\widehat{\boldsymbol{\Phi}}=(\hat{\boldsymbol{\varphi}}_1,\ldots,\hat{\boldsymbol{\varphi}}_{\hat{r}})$ and $\hat{\boldsymbol{\varphi}}_j$ is the eigenvector corresponding to eigenvalue $\hat{\varphi}_j$ for $j=1,\ldots,\hat{r}$ 
\vspace{2mm}

Transform data using the projection matrix 
\vspace{2mm}
Divide the sample into $K$ folds, and randomly assign subject $i$ and its time series to fold $k$ such that the estimation sample is $W_k$ with size $|N_k|$, and the complementary sample $W_k^c$.

        

\For{$k\leftarrow 1$ \KwTo $K$}{
        Predict $\widehat{\bm{\eta}}_k$ using base learners on complementary data $\mathcal{W}_k^c$ .\\
        Obtain orthogonalised residuals using $\widehat{\bm{\eta}}_k$, and construct Neyman orthogonal score function.\\
        Solve empirical moment condition for $\theta_k$ using estimation sample $W_k$.
    }
    \eIf{DML1 Procedure}
    {
        Average the intermediate estimates of $\{\widehat{\theta}_k\}_{k = 1}^K$ and model RMSE, compute finite-sample variance-covariance matrix $\widehat{\bm{\Sigma}}$, and  RMSE~of~$\{l,m,g\}$. 
    }{ \If{DML2 Procedure}
        {
        Obtain final estimate of $\{\widehat{\theta}_k\}_{k = 1}^K$ by aggregating intermediate results and model RMSE, compute finite-sample variance-covariance matrix $\widehat{\bm{\Sigma}}$, and  RMSE~of~$\{l,m,g\}$.
        }
    }
\end{algorithm}


\section{Monte Carlo Simulation}\label{sec:mcsims}
\setcounter{equation}{0}

We assess the finite-sample performance of the proposed panel DML-IFE estimator through Monte Carlo simulations applied to data generated under three nuisance function-specifications of increasing degrees of non-linearity. Throughout, we benchmark against the linear IFE estimator of \citet{bai2009panel}, the canonical estimator for the IFE setting when the observed covariates are assumed to enter linearly and no high-dimensional adjustment is required.

\subsection{Data-generating Process and Estimators}

Consider the partially linear panel regression with interactive fixed effects (PLPR-IFE),
\begin{align}
Y_{it} &= V_{it}\,\theta_0 + l_0(\mathbf{X}_{it}) + \vector{\lambda}^\top_i \vector{f}_t + U_{it}, \label{eqn:y_mc}\\
V_{it} &=  D_{it} - m_0(\mathbf{X}_{it}) - \vector{\phi}_i^\top \vector{f}_t, \label{eqn:g_mc} \\
\mathbf{X}_{it} &= \boldsymbol{\Gamma}_i \vector{f}_t + \boldsymbol{E}_{it}, \label{eqn:x_mc}
\end{align}
with structural parameter $\theta_0=1$; mutually independent idiosyncratic errors $U_{it}, V_{it}, E_{it,j}\sim N(0,1)$. For each of the $r=2$ factors, $\phi_i,\lambda_i\sim N(0,1)$ and $\Gamma_i\sim N(0,9)$; and factors $f_t\sim N(0,1)$. The covariate dimension varies over $p\in\{50,100,300\}$, of which only the first two covariates are relevant, so that the sparsity condition $s\ll p$ holds with $s=2$.

The nuisance functions $(l_0,m_0)$ are generated under three designs of increasing complexity. The linear design (DGP1) corresponds to the specification most commonly assumed in empirical practice,
\begin{align*}
l_0(\mathbf{X}_{it}) &= a_1 X_{it,1} + a_2 X_{it,2}, \\
m_0(\mathbf{X}_{it}) &= b_1 X_{it,1} + b_2 X_{it,2}.
\end{align*}
The nonlinear and smooth design (DGP2) introduces non-smoothness in the relevant covariates,
\begin{align*}
l_0(\mathbf{X}_{it}) &= a_1\max(X_{it,1},0) + a_2 |X_{it,2}|, \\
m_0(\mathbf{X}_{it}) &= b_1\,\one\{X_{it,1}>0\} + b_2 \log(1+|X_{it,2}|).
\end{align*}
The nonlinear and discontinuous design (DGP3), following \citet{clarke2025double}, combines interactions with indicator transformations,
\begin{align*}
l_0(\mathbf{X}_{it}) &= a_1(X_{it,1}\cdot X_{it,2}) + a_2(X_{it,2}\cdot \one\{X_{it,2}>0\}), \\
m_0(\mathbf{X}_{it}) &= b_1(X_{it,1}\cdot \one\{X_{it,1}>0\}) + b_2(X_{it,1}\cdot X_{it,2}),
\end{align*}
with $a_1=b_2=0.25$ and $a_2=b_1=0.5$ throughout, and $\one\{\cdot\}$ the indicator operator.

We generate data under PLPR-IFE model~\eqref{eqn:y_mc}-\eqref{eqn:x_mc} using each of the three different data generating processes (DGPs) defined above.  The baseline estimator against which the DML-IFE procedure is compared is the linear IFE estimator of \citet{bai2009panel}.  The three versions of the DML-IFE estimator differ by the choice of first-stage learner:\ Lasso; gradient boosting (Boosting); and feed-forward neural networks (NNet).\footnote{Monte Carlo simulations are executed in \textsc{R} on the Ceres HPC facility at the University of Essex. IFE standard errors are cluster-robust and computed via the \texttt{xtife} package \citep{xtife}.}

For the linear IFE estimator and DML-IFE with Boosting and NNet, the conditioning set comprises the $p$ raw covariates; the linear estimator models all $p$ covariates linearly, while DML with Boosting and NNet learn the true dimensionality and (potentially non-linear) functional form of $(l_0,m_0)$. 
Lasso, on the other hand, requires the analyst to specify an extended dictionary of polynomials and pairwise interactions involving the $p$ covariates, from which it selects the the combination of terms that best approximate the true functions. (Note that we deliberately exclude the specific nonlinear transformations defining DGP2 and DGP3 from this dictionary to be able to assess how effectively Lasso works in practice.) 

Each of the learners involves hyperparameters that dictate its performance and which must be set by the analyst.  The Lasso penalty is selected by cross-validation at minimum cross-validation error. Boosting and NNet hyperparameters are tuned by a random search as described in Online Appendix~\ref{sec:tuning}.\footnote{The choice of random search over grid search follows \citet{bergstra2012}, who show that randomly chosen trials are more efficient for hyperparameter optimisation than grids.}

\subsection{Monte Carlo Results}

Tables~\ref{tab:dgp1_n20}-\ref{tab:dgp3_n20} report results from $R=100$ Monte Carlo replications for cross-sectional sample size $N=20$, time dimension $T\in\{30,50,100\}$ (columns), and covariate dimension $p\in\{50,100,300\}$ (panels), with one column per estimator.\footnote{The combination $(N,T,p)$ is calibrated to the empirical application in Section~\ref{sec:empirical}, for which $(N=29,T=72,p=89)$. Results for larger cross-sectional samples $N\in\{50,100,500\}$ are reported in Appendix~\ref{sec:app_tabs}.} 
We report the following metrics: 
The Monte Carlo root mean squared error of the structural estimator,
\begin{equation*}
\text{MC RMSE} = \sqrt{\frac{1}{R}\sum_{r=1}^{R}\bigl(\widehat\theta_r - \theta_0\bigr)^2}.
\end{equation*}
The empirical standard deviation of $\widehat\theta$ across replications
(SD) and the average analytic standard error (SE). Recalling that all variables and nuisance functions have been transformed using $\hat{\Pi}$ (e.g.\ $\tilde{Y}_{i}=\hat{\Pi}Y_{it}$ and $\hat{l}(\boldsymbol{X}_{it})=\widehat{\Pi}{l_0(\boldsymbol{X}_{it})}$), we have
\begin{equation*}
\text{Model RMSE} = \frac{1}{R}\sum_{r=1}^{R}\sqrt{\frac{1}{NT}\sum_{i=1}^{N}\sum_{t=1}^{T}\Bigl(\tilde{Y}_{it} - \hat{l}_r(\mathbf{X}_{it}) - \bigl(\tilde{D}_{it} - \widehat{m}_r(\mathbf{X}_{it})\bigr)\widehat\theta_r\Bigr)^2},
\end{equation*}
and the RMSEs of the individual nuisance-function estimates,
\begin{equation*}
\text{RMSE}_{\widehat g} = \frac{1}{R}\sum_{r=1}^{R}\sqrt{\frac{1}{NT}\sum_{i=1}^{N}\sum_{t=1}^{T}\bigl(\widehat{g}_r(\mathbf{X}_{it}) - g_0(\mathbf{X}_{it})\bigr)^2},
\qquad g\in\{l,m\}.
\end{equation*}
where $g=\{l,m\}$ indexes the nuisance function.  Performance is discussed along these dimensions with particular attention to the interaction between DGP complexity, $T$,~and~$p$. We report the \emph{feasible} bias of the estimator of the structural parameter, and the \emph{infeasible} bias with $\widehat{\Pi}$ replaced by $\Pi_0$.

\begin{table}[t!]
\centering
\caption{Monte Carlo Results: DGP 1, $N = 20$}
\label{tab:dgp1_n20}
\scalebox{.65}{
\begin{threeparttable}
\begin{tabular}{lcccccccccccc}
\vspace{-3mm}\\
\hline\hline
\vspace{-3mm}\\
 & \multicolumn{4}{c}{$T=30$} & \multicolumn{4}{c}{$T=50$} & \multicolumn{4}{c}{$T=100$} \\
\cmidrule(lr){2-5}\cmidrule(lr){6-9}\cmidrule(lr){10-13}
 & IFE & Lasso & Boosting & NNet &IFE & Lasso & Boosting &  NNet &IFE & Lasso & Boosting &  NNet\\
\hline
\vspace{-3mm}\\
\multicolumn{13}{c}{\emph{Panel A: p = 50}}\\
Bias & 0.0280 & 0.0382 & -0.0814 & -0.2094 & 0.0110 & 0.0316 & -0.0469 & -0.1220 & 0.0042 & 0.0324 & -0.0133 & -0.0805 \\
Infeasible bias & -0.0020 & -0.0004 & -0.1102 & -0.1921 & 0.0027 & 0.0055 & -0.0738 & -0.1438 & -0.0010 & 0.0038 & -0.0414 & -0.0985 \\
MC RMSE & 0.0058 & 0.0035 & 0.0093 & 0.0691 & 0.0017 & 0.0019 & 0.0041 & 0.0247 & 0.0008 & 0.0017 & 0.0011 & 0.0287 \\
SD & 0.0711 & 0.0452 & 0.0520 & 0.1596 & 0.0393 & 0.0303 & 0.0437 & 0.0995 & 0.0274 & 0.0266 & 0.0305 & 0.1499 \\
SE & 1.6615 & 0.0408 & 0.0450 & 0.0612 & 1.0987 & 0.0314 & 0.0334 & 0.0404 & 0.8083 & 0.0224 & 0.0237 & 0.0381 \\
Model RMSE &   & 5.5280 & 6.6704 & 7.8778 &   & 7.0997 & 8.3205 & 9.1921 &   & 10.1368 & 11.1555 & 14.7760 \\
RMSE$_l$ &   & 1.4478 & 1.5905 & 2.1192 &   & 1.4477 & 1.5717 & 1.7549 &   & 1.4618 & 1.5368 & 2.1627 \\
RMSE$_m$ &   & 1.0091 & 1.1077 & 1.5143 &   & 1.0091 & 1.0957 & 1.3022 &   & 1.0174 & 1.0680 & 1.2529 \\
\multicolumn{13}{c}{\emph{Panel B: p = 100}}\\
Bias & 0.0618 & 0.0210 & -0.0986 & -0.4452 & 0.0224 & 0.0212 & -0.0702 & -0.3742 & 0.0082 & 0.0165 & -0.0355 & -0.2169 \\
Infeasible bias & -0.0022 & 0.0013 & -0.1102 & -0.4165 & -0.0017 & 0.0041 & -0.0948 & -0.3974 & -0.0021 & 0.0021 & -0.0482 & -0.2375 \\
MC RMSE & 0.0080 & 0.0023 & 0.0127 & 0.2443 & 0.0028 & 0.0015 & 0.0072 & 0.2043 & 0.0007 & 0.0007 & 0.0020 & 0.1159 \\
SD & 0.0647 & 0.0435 & 0.0552 & 0.2159 & 0.0487 & 0.0333 & 0.0475 & 0.2548 & 0.0249 & 0.0214 & 0.0270 & 0.2638 \\
SE & 1.7965 & 0.0403 & 0.0456 & 0.0714 & 1.3588 & 0.0311 & 0.0322 & 0.0581 & 0.8588 & 0.0217 & 0.0235 & 0.0492 \\
Model RMSE &   & 5.4271 & 6.7079 & 10.4108 &   & 7.0428 & 8.3198 & 24.0787 &   & 10.0530 & 11.2931 & 20.0689 \\
RMSE$_l$ &   & 1.4257 & 1.5725 & 2.7521 &   & 1.4339 & 1.5517 & 2.6462 &   & 1.4343 & 1.5178 & 2.9350 \\
RMSE$_m$ &   & 1.0030 & 1.0924 & 2.6313 &   & 1.0067 & 1.0863 & 5.4574 &   & 1.0080 & 1.0599 & 4.6694 \\
\multicolumn{13}{c}{\emph{Panel C: p = 300}}\\
Bias & 0.0559 & 0.0287 & -0.1112 & -0.1786 & 0.0493 & 0.0161 & -0.1048 & -0.4069 & 0.0291 & 0.0143 & -0.0590 & -0.6655 \\
Infeasible bias & -0.0070 & 0.0151 & -0.1162 & -0.1721 & 0.0040 & 0.0095 & -0.1108 & -0.4594 & -0.0013 & 0.0022 & -0.0649 & -0.6834 \\
MC RMSE & 0.0128 & 0.0028 & 0.0145 & 0.0416 & 0.0049 & 0.0013 & 0.0128 & 0.1926 & 0.0020 & 0.0006 & 0.0041 & 0.5068 \\
SD & 0.0989 & 0.0444 & 0.0462 & 0.0988 & 0.0501 & 0.0320 & 0.0428 & 0.1652 & 0.0335 & 0.0209 & 0.0254 & 0.2540 \\
SE & 3.4808 & 0.0413 & 0.0463 & 0.0483 & 1.2036 & 0.0301 & 0.0339 & 0.0482 & 0.7292 & 0.0216 & 0.0226 & 0.0363 \\
Model RMSE &   & 5.4860 & 6.6484 & 9.8610 &   & 7.1193 & 8.4675 & 15.4579 &   & 10.0088 & 11.3297 & 36.0501 \\
RMSE$_l$ &   & 1.4379 & 1.5496 & 2.2164 &   & 1.4280 & 1.5485 & 2.4755 &   & 1.4275 & 1.5149 & 3.5796 \\
RMSE$_m$ &   & 1.0088 & 1.0819 & 1.5496 &   & 1.0049 & 1.0883 & 2.0914 &   & 1.0058 & 1.0631 & 5.5722 \\
\hline
\end{tabular}
\begin{tablenotes}[para,flushleft]
\textbf{Note:} The figures in the table are average values and frequencies over 100 replications by estimation method, i.e.,  IFE by \citet{bai2009panel}, and our panel DML-IFE with Lasso, gradient boosting, and neural network. The true structural parameter $\theta$ is 1.  Standard errors in parenthesis are clustered at the firm level.  Additional details on panel DML-IFE estimation: cross-fitting with 5 folds. Lasso hyperparameter is selected from the model with minimum cross-validated error; gradient boosting and neural network hyperparameters are tuned with random search. 
\end{tablenotes}
\end{threeparttable}}
\end{table}

\paragraph{Linear design (DGP1).} Under the linear design (Table~\ref{tab:dgp1_n20}), as expected, the feasible IFE estimator performs comparably to DML-Lasso and DML-Boosting in terms of bias when $p$ is small or $T$ large, with the feasible bias decreasing monotonically in $T$ for fixed $p$. The infeasible IFE bias is negligible across all $(T,p)$, as expected, confirming correct specification under linearity; the feasible estimator does not achieve the oracle's accuracy at $N=20$ because of residual sampling variability in the estimated factor structure. The $\mathrm{SE}/\mathrm{SD}$ ratio exceeds unity by one to two orders of magnitude, making IFE inference severely conservative, primarily because this estimator does not select out irrelevant covariates (i.e.\ those with no effect on the mean).

Among the three DML-IFE specifications, DML-Lasso achieves the lowest or near-lowest bias and MC RMSE across all $(T,p)$. Its infeasible bias is close to zero, confirming that the regularised linear approximation introduces no systematic distortion in the linear DGP. The SE and SD are close in magnitude, indicating that inference will be accurate. DML-Boosting underestimates $\theta_0$, producing a persistent negative bias ranging from $-0.013$ to $-0.111$, with both feasible bias and MC RMSE declining in $T$. The infeasible bias is also negative, close in magnitude to the feasible bias, and approximately invariant to $T$. This behaviour indicates systematic learner misspecification; specifically, boosting cannot match a linear signal exactly even when the factor structure is known. Standard errors accurately capture the standard deviation of the sampling distribution.  However, DML-NNet exhibits unstable results:\ although the bias shrinks with $T$ at $p=50$, this is not the case for larger $p$ and its performance deteriorates. The infeasible bias mirrors the feasible bias in both sign and magnitude, confirming that the bias is due to the neural network failing to approximate the simple linear signal in high dimensions at $N=20$. The MC RMSE does not decrease monotonically in $(T,p)$, and Model RMSE reaches $36.1$ at $(T=100,p=300)$, indicating severe misspecification of the first stage in high-dimensional linear environments.


\begin{table}[t!]
\centering
\caption{Monte Carlo Results: DGP 2, $N = 20$}
\label{tab:dgp2_n20}
\scalebox{.6}{
\begin{threeparttable}
\begin{tabular}{lcccccccccccc}
\vspace{-3mm}\\
\hline\hline
\vspace{-3mm}\\
 & \multicolumn{4}{c}{$T=30$} & \multicolumn{4}{c}{$T=50$} & \multicolumn{4}{c}{$T=100$} \\
\cmidrule(lr){2-5}\cmidrule(lr){6-9}\cmidrule(lr){10-13}
 & IFE & Lasso & Boosting & NNet &IFE & Lasso & Boosting &  NNet &IFE & Lasso & Boosting &  NNet\\
\hline
\multicolumn{13}{c}{\emph{Panel A: p = 50}}\\
Bias & 0.0400 & 0.0873 & -0.0497 & -0.1760 & 0.0279 & 0.0796 & -0.0109 & -0.0017 & 0.0239 & 0.0750 & 0.0261 & -0.0401 \\
Infeasible bias & 0.1593 & 0.0211 & -0.1024 & -0.1730 & 0.1646 & 0.0186 & -0.0789 & -0.1793 & 0.1580 & 0.0128 & -0.0376 & -0.0148 \\
MC RMSE & 0.0047 & 0.0101 & 0.0047 & 0.0741 & 0.0021 & 0.0075 & 0.0023 & 0.4096 & 0.0012 & 0.0066 & 0.0017 & 0.0224 \\
SD & 0.0557 & 0.0498 & 0.0479 & 0.2086 & 0.0365 & 0.0343 & 0.0468 & 0.6432 & 0.0256 & 0.0319 & 0.0323 & 0.1448 \\
SE & 3.8549 & 0.0406 & 0.0450 & 0.1037 & 2.6705 & 0.0327 & 0.0352 & 0.1376 & 1.9452 & 0.0246 & 0.0255 & 0.0453 \\
Model RMSE &   & 5.7016 & 6.9169 & 21.5463 &   & 7.3137 & 8.6608 & 12.9021 &   & 10.4290 & 11.2989 & 13.7719 \\
RMSE$_l$ &   & 1.5386 & 1.6630 & 3.8614 &   & 1.5347 & 1.6542 & 4.9015 &   & 1.5404 & 1.6047 & 3.1034 \\
RMSE$_m$ &   & 1.0516 & 1.1407 & 2.1586 &   & 1.0490 & 1.1285 & 1.2135 &   & 1.0519 & 1.0985 & 1.6671 \\
\multicolumn{13}{c}{\emph{Panel B: p = 100}}\\
Bias & 0.0890 & 0.0562 & -0.0808 & -0.4170 & 0.0345 & 0.0514 & -0.0643 & -0.2166 & 0.0241 & 0.0431 & -0.0187 & -0.3265 \\
Infeasible bias & 0.1569 & 0.0227 & -0.1046 & -0.3710 & 0.1576 & 0.0180 & -0.0990 & -0.2710 & 0.1574 & 0.0113 & -0.0490 & -0.2599 \\
MC RMSE & 0.0146 & 0.0050 & 0.0087 & 0.2679 & 0.0033 & 0.0038 & 0.0059 & 0.4227 & 0.0011 & 0.0024 & 0.0012 & 0.6470 \\
SD & 0.0821 & 0.0433 & 0.0473 & 0.3081 & 0.0462 & 0.0345 & 0.0419 & 0.6161 & 0.0237 & 0.0225 & 0.0294 & 0.7388 \\
SE & 4.0009 & 0.0402 & 0.0454 & 0.0972 & 3.2303 & 0.0314 & 0.0339 & 0.1613 & 2.1318 & 0.0224 & 0.0236 & 0.1936 \\
Model RMSE &   & 5.5597 & 6.8181 & 15.6625 &   & 7.1985 & 8.5083 & 23.6896 &   & 10.2418 & 11.3974 & 61.3228 \\
RMSE$_l$ &   & 1.4917 & 1.6059 & 4.3691 &   & 1.4916 & 1.5904 & 7.1297 &   & 1.4854 & 1.5504 & 12.5741 \\
RMSE$_m$ &   & 1.0323 & 1.1074 & 2.8552 &   & 1.0343 & 1.1052 & 4.6031 &   & 1.0318 & 1.0763 & 5.3041 \\
\multicolumn{13}{c}{\emph{Panel C: p = 300}}\\
Bias & 0.1421 & 0.0542 & -0.0941 & -0.2550 & 0.1078 & 0.0335 & -0.0991 & -0.4376 & 0.0524 & 0.0267 & -0.0541 & -0.6286 \\
Infeasible bias & 0.1517 & 0.0375 & -0.1115 & -0.2028 & 0.1591 & 0.0239 & -0.1131 & -0.4263 & 0.1590 & 0.0117 & -0.0628 & -0.5929 \\
MC RMSE & 0.0287 & 0.0049 & 0.0113 & 0.0982 & 0.0145 & 0.0022 & 0.0116 & 0.2469 & 0.0041 & 0.0013 & 0.0036 & 0.4770 \\
SD & 0.0926 & 0.0445 & 0.0498 & 0.1831 & 0.0542 & 0.0336 & 0.0421 & 0.2366 & 0.0363 & 0.0243 & 0.0263 & 0.2876 \\
SE & 4.6798 & 0.0402 & 0.0443 & 0.0523 & 2.4358 & 0.0305 & 0.0341 & 0.0640 & 1.9177 & 0.0214 & 0.0225 & 0.0797 \\
Model RMSE &   & 5.4954 & 6.7153 & 10.8827 &   & 7.1483 & 8.5024 & 17.1556 &   & 10.1797 & 11.4030 & 64.7673 \\
RMSE$_l$ &   & 1.4746 & 1.5789 & 2.3925 &   & 1.4650 & 1.5684 & 3.3749 &   & 1.4607 & 1.5307 & 8.2246 \\
RMSE$_m$ &   & 1.0219 & 1.0965 & 1.9630 &   & 1.0227 & 1.1015 & 2.2720 &   & 1.0223 & 1.0748 & 5.8153 \\
\hline
\end{tabular}
\begin{tablenotes}[para,flushleft]
\textbf{Note:} The figures in the table are average values and frequencies over 100 replications by estimation method, i.e.,  IFE by \citet{bai2009panel}, and our panel DML-IFE with Lasso, gradient boosting, and neural network. The true structural parameter $\theta$ is 1.  Standard errors in parenthesis are clustered at the firm level.  Additional details on panel DML-IFE estimation: cross-fitting with 5 folds. Lasso hyperparameter is selected from the model with minimum cross-validated error; gradient boosting and neural network hyperparameters are tuned with random search. 
\end{tablenotes}
\end{threeparttable}}
\end{table}

\paragraph{Nonlinear smooth design (DGP2).} Under the nonlinear and smooth design (Table~\ref{tab:dgp2_n20}), the IFE estimator is competitive with DML-IFE in bias only at $p=50$ but (as expected) never in precision. The feasible IFE bias is, as expected, positive and small at low $p$ but does not decrease with either $T$ or $p$ in contrast to DML-IFE. The infeasible bias stabilises around $0.158$ across $(T,p)$, but its absolute values are larger than its feasible bias. This is difficult to interpret but possibly reveals that the systematic misspecification of the covariates' functional form is slightly offset by the estimation error of the factor structure. 
Just as under DGP1, $\mathrm{SE}\gg\mathrm{SD}$, IFE  inference is severely conservative. MC RMSE declines monotonically in $T$ for fixed $p$, indicating $T$-convergence to a non-zero pseudo-true limit.

Among the DML-IFE specifications, DML-Lasso delivers the lowest bias and MC RMSE across $(T,p)$, exploiting the regularised expansion over the extended dictionary to approximate the smooth nonlinear nuisance functions. The infeasible bias is close to zero throughout, confirming that the dictionary captures the relevant functional features, and the gap between the feasible and infeasible biases shrinks with $T$. The $\mathrm{SE}/\mathrm{SD}$ ratio remains close to unity, and Model RMSE is moderate, with $\mathrm{RMSE}_{\widehat l}$ and $\mathrm{RMSE}_{\widehat m}$ of comparable magnitude.

DML-Boosting displays biases which are comparable to, and sometimes smaller than, those of DML-Lasso, while systematically underestimating $\theta_0$. The infeasible bias is non-negligible and approximately invariant to $T$, indicating that gradient boosting incurs a systematic approximation error against the smooth nuisance functions even under the oracle-not finite-sample noise. The relation between the feasible and infeasible biases again potentially reflects partial offsetting between learner misspecification and an the contribution of the factor-estimation in the opposite direction.\footnote{Block-$k$-fold cross-fitting is performed at the unit level, so the test sample within each fold is quite small relative to standard cross-sectional DML applications; this magnifies finite-sample variability in the first stage, but as the unfeasibel bias is stable in $T$ this variability does not by itself explain the observed bias.} Model RMSE exceeds that of DML-Lasso, being consistent with the relative disadvantage of tree-based learners in smooth environments. The $\mathrm{SE}/\mathrm{SD}$ ratio remains close to unity, so inference remains reliable notwithstanding the structural bias.

In contrast, DML-NNet estimates deteriorate substantially as $p$ grows, underestimating $\theta_0$ by a margin larger than DML-Boosting. The infeasible bias is of comparable magnitude to the feasible bias and does not attenuate with $T$, identifying the dominant source as the inability of the neural network to recover $(l_0,m_0)$ under the true factor structure at $N=20$. Model RMSE reaches $64.77$ at $(T=100,p=300)$, roughly six times that of DML-Lasso and more than five times that of DML-Boosting; MC RMSE is the highest across estimators and does not decrease uniformly in $(T,p)$, indicating the breakdown of $\sqrt{NT}$-consistency in this configuration. The pattern is consistent with the documented difficulty of training neural networks in panels with small $N$ and high-dimensional covariates. The good news for practice is that using an ensemble strategy based on choosing the learner performing best in terms of RMSE would always exclude DML-NNET.



\begin{table}[t!]
\centering
\caption{Monte Carlo Results: DGP 3, $N = 20$}
\label{tab:dgp3_n20}
\scalebox{.6}{
\begin{threeparttable}
\begin{tabular}{lcccccccccccc}
\vspace{-3mm}\\
\hline\hline
\vspace{-3mm}\\
 & \multicolumn{4}{c}{$T=30$} & \multicolumn{4}{c}{$T=50$} & \multicolumn{4}{c}{$T=100$} \\
\cmidrule(lr){2-5}\cmidrule(lr){6-9}\cmidrule(lr){10-13}
 & IFE & Lasso & Boosting & NNet &IFE & Lasso & Boosting &  NNet &IFE & Lasso & Boosting &  NNet\\
\hline
\multicolumn{13}{c}{\emph{Panel A: p = 50}}\\
Bias & 0.2446 & 0.1261 & 0.0389 & -0.1506 & 0.2185 & 0.1054 & 0.0168 & -0.0380 & 0.2066 & 0.1045 & 0.0384 & 0.0050 \\
Infeasible bias & 0.1994 & 0.0503 & 0.0013 & -0.1263 & 0.2058 & 0.0265 & -0.0177 & -0.1240 & 0.2033 & 0.0155 & -0.0491 & -0.0547 \\
MC RMSE & 0.0684 & 0.0183 & 0.0047 & 0.3993 & 0.0514 & 0.0129 & 0.0027 & 0.0160 & 0.0435 & 0.0120 & 0.0043 & 0.0016 \\
SD & 0.0930 & 0.0490 & 0.0564 & 0.6168 & 0.0607 & 0.0428 & 0.0493 & 0.1214 & 0.0283 & 0.0333 & 0.0538 & 0.0398 \\
SE & 2.0938 & 0.0440 & 0.0485 & 0.1131 & 1.5382 & 0.0335 & 0.0391 & 0.0523 & 1.1169 & 0.0265 & 0.0279 & 0.0305 \\
Model RMSE &   & 5.6882 & 7.4114 & 22.5806 &   & 7.4205 & 9.5393 & 12.0920 &   & 10.5316 & 12.8444 & 12.8755 \\
RMSE$_l$ &   & 1.5860 & 1.8409 & 4.1396 &   & 1.5674 & 1.8077 & 2.1802 &   & 1.5718 & 1.7567 & 1.7253 \\
RMSE$_m$ &   & 1.0555 & 1.1948 & 1.9199 &   & 1.0503 & 1.1830 & 1.6022 &   & 1.0559 & 1.1548 & 1.1483 \\
\multicolumn{13}{c}{\emph{Panel B: p = 100}}\\
Bias & 0.2934 & 0.0834 & 0.0322 & -0.3571 & 0.2414 & 0.0747 & 0.0089 & -0.3063 & 0.2152 & 0.0606 & -0.0273 & -0.1923 \\
Infeasible bias & 0.2041 & 0.0449 & 0.0098 & -0.2568 & 0.1982 & 0.0338 & -0.0107 & -0.2392 & 0.2056 & 0.0169 & -0.0541 & -0.1951 \\
MC RMSE & 0.0924 & 0.0088 & 0.0042 & 0.2660 & 0.0625 & 0.0069 & 0.0025 & 0.3299 & 0.0479 & 0.0044 & 0.0028 & 0.3586 \\
SD & 0.0797 & 0.0431 & 0.0570 & 0.3740 & 0.0655 & 0.0371 & 0.0499 & 0.4883 & 0.0402 & 0.0279 & 0.0450 & 0.5700 \\
SE & 2.1478 & 0.0409 & 0.0466 & 0.1348 & 1.7744 & 0.0328 & 0.0367 & 0.1568 & 1.2249 & 0.0232 & 0.0269 & 0.1770 \\
Model RMSE &   & 5.5705 & 7.2872 & 25.2600 &   & 7.2780 & 9.4796 & 27.5990 &   & 10.3370 & 13.1072 & 44.8228 \\
RMSE$_l$ &   & 1.5237 & 1.8200 & 6.0491 &   & 1.5227 & 1.7872 & 8.4983 &   & 1.5043 & 1.7131 & 9.2038 \\
RMSE$_m$ &   & 1.0398 & 1.1778 & 3.2384 &   & 1.0367 & 1.1699 & 2.9930 &   & 1.0308 & 1.1386 & 4.0032 \\
\multicolumn{13}{c}{\emph{Panel C: p = 300}}\\
Bias & 0.2947 & 0.0630 & 0.0442 & -0.0573 & 0.2743 & 0.0438 & 0.0142 & -0.3251 & 0.2427 & 0.0326 & -0.0243 & -0.6270 \\
Infeasible bias & 0.2027 & 0.0505 & 0.0253 & -0.0823 & 0.2086 & 0.0296 & 0.0037 & -0.3529 & 0.2059 & 0.0169 & -0.0310 & -0.5770 \\
MC RMSE & 0.0974 & 0.0058 & 0.0047 & 0.0770 & 0.0775 & 0.0029 & 0.0023 & 0.1252 & 0.0600 & 0.0016 & 0.0022 & 0.4731 \\
SD & 0.1034 & 0.0428 & 0.0528 & 0.2728 & 0.0476 & 0.0319 & 0.0456 & 0.1403 & 0.0339 & 0.0242 & 0.0401 & 0.2842 \\
SE & 3.5354 & 0.0415 & 0.0487 & 0.0692 & 1.4598 & 0.0317 & 0.0366 & 0.0579 & 0.9695 & 0.0215 & 0.0267 & 0.0618 \\
Model RMSE &   & 5.5537 & 7.3579 & 14.6026 &   & 7.1941 & 9.4121 & 19.2559 &   & 10.1663 & 13.3037 & 42.4584 \\
RMSE$_l$ &   & 1.4964 & 1.8263 & 3.0900 &   & 1.4756 & 1.7965 & 3.2833 &   & 1.4626 & 1.7329 & 5.5272 \\
RMSE$_m$ &   & 1.0273 & 1.1770 & 1.8837 &   & 1.0231 & 1.1726 & 1.8998 &   & 1.0185 & 1.1404 & 4.7788 \\
\hline
\end{tabular}
\begin{tablenotes}[para,flushleft]
\textbf{Note:} The figures in the table are average values and frequencies over 100 replications by estimation method, i.e.,  IFE by \citet{bai2009panel}, and our panel DML-IFE with Lasso, gradient boosting, and neural network. The true structural parameter $\theta$ is 1.  Standard errors in parenthesis are clustered at the firm level.  Additional details on panel DML-IFE estimation: cross-fitting with 5 folds. Lasso hyperparameter is selected from the model with minimum cross-validated error; gradient boosting and neural network hyperparameters are tuned with random search. 
\end{tablenotes}
\end{threeparttable}}
\end{table}

\paragraph{Discontinuous design (DGP3).} The discontinuous design (Table~\ref{tab:dgp3_n20}) constitutes the most challenging environment for all estimators. The IFE estimator produces the largest biases, with the feasible bias exceeding $0.20$ in nearly every $(T,p)$ configuration and showing no meaningful attenuation in $T$. The infeasible bias remains around $0.20$ throughout, identifying the source of bias as systematic misspecification. Specifically, an estimator that imposes a linear parametrisation in the covariates cannot approximate discontinuous nuisance functions $(l_0,m_0)$. The closeness of the feasible and infeasible biases further indicates that the contribution of factor-estimation error is small relative to misspecification under this DGP. Standard errors are extremely large relative to both the empirical SD and the corresponding SEs of the DML-IFE estimators, signalling unreliable IFE inference due to small sample size and failure to select out irrelevant covariates.

Among the DML-IFE specifications, DML-Lasso exhibits a persistent positive bias of $0.03$-$0.13$ that decreases in $T$ for fixed $p$. The learner performs well in recovering the causal effect, suggesting that a sufficiently rich dictionary of polynomial and interaction terms can partially approximate the discontinuity even without it being explicitly specified. The decomposition of the bias is informative here. At $p\le 100$, the feasible bias is almost double the infeasible bias with both terms of the same sign, so factor-estimation error contributes roughly as much as learner misspecification to the total bias. As $p$ grows, the feasible and infeasible biases converge, indicating that the dictionary expansion contains the relevant interactions and reduces the factor-estimation error transmitted through the within-projection. Inference appears reliable, with $\mathrm{SE}/\mathrm{SD}\approx 1$ across all $(T,p)$, confirming that the cross-fitting procedure delivers accurate variance estimation under functional-form misspecification. MC RMSE remains low and decreases in $T$ and $p$, consistent with $\sqrt{NT}$-convergence. Model RMSE is larger than under DGP1 and DGP2, reflecting the more complex learning task, with $\mathrm{RMSE}_{\widehat l}>\mathrm{RMSE}_{\widehat m}$ throughout.

DML-Boosting achieves the most favourable balance of bias and MC RMSE under DGP3, reflecting the known advantage of tree-based learners in approximating piecewise-constant and discontinuous functions. The feasible bias is closest to zero across most $(T,p)$ combinations and declines monotonically in $T$. At low $T$, the feasible bias is larger than the infeasible bias with both of the same sign, indicating a non-negligible factor-estimation contribution. The two converge as $T$ and $p$ grow, identifying gradient-boosting approximation error as the dominant residual source of bias in larger samples. Standard errors are broadly aligned with empirical standard deviations; in larger and higher-dimensional configurations, $\mathrm{SE}$ falls slightly below $\mathrm{SD}$, implying that the cluster-robust standard errors moderately understate dispersion and may lead to mild over-rejection in those cells.

DML-NNet again exhibits severe instability. At $p=50$, the estimator produces moderate negative biases that shrink with $T$, suggesting partial approximation of the discontinuous signal when the feature space is manageable. However, as $p$ grows, its performance deteriorates:\ the infeasible bias is large and negative, identifying the failure as the inability of the network to approximate discontinuous nuisance functions in high dimensions at small $N$. MC RMSE does not decrease uniformly with $T$ and $p$, signalling the breakdown of $\sqrt{NT}$-consistency, and DML-NNet produces the highest Model RMSE and nuisance-function RMSEs in the simulation. The pattern is consistent with the theoretical observation that neural networks require substantially larger samples to resolve discontinuities than smooth functions, a requirement that is not met at $N=20$ when $p$ is of comparable or larger magnitude.

\paragraph{Large-$N$ results.} The large-$N$ results reported in Appendix~\ref{sec:app_tabs} (for $N\in\{50,100,500\}$) confirm two patterns. Increasing $N$ reduces the variance of every estimator but reduces bias only for the flexible learners, and even then conditional on the DGP. The IFE estimator remains structurally biased outside the linear case regardless of $N$, consistent with the diagnostic that the infeasible bias stabilises at a non-zero value under DGP2 and DGP3.

Taken together, the simulation evidence supports four conclusions. First, DML-Lasso delivers the most consistent finite-sample performance across DGPs, time and covariate dimensions, making it the preferred choice under uncertainty about the functional form of $(l_0,m_0)$, provided that a sufficiently rich dictionary is supplied; its bias is reduced primarily by larger $T$ rather than larger $N$, so additional cross-sectional units yield variance reductions but limited bias improvements. Second, DML-Boosting offers a meaningful advantage in settings with  discontinuities, at the cost of elevated Model RMSE in nonlinear smooth environments; unlike Lasso, it benefits substantially from larger $N$ in both bias and variance. Third, DML-NNet is the least stable of the learners considered, with results driven by occasional extreme estimates; larger $N$ improves its average performance but it is always outperformed by Lasso  and Boosting in terms of bias, and its standard errors underestimate the standard deviation severely.  Fourth, the IFE estimator is unbiased and well-behaved in the correctly-specified linear case (DGP1), where it converges to $\theta_0$ as $N$ increases, but elsewhere it is practically uninformative because of its inflated standard errors and a structural bias that additional cross-sectional units cannot remove.\
\section{Empirical Application} \label{sec:empirical} 
We illustrate the applicability of our method to an empirical exercise to study the effect of several firm characteristics on monthly excess stock return. This application is based on the data and empirical analysis conducted in \citet{rucker2025}.\footnote{For the construction of the dataset, we referred to the replication package available on the authors' GitHub repository: \url{https://github.com/RueckerM/hdcce-ReplicationFiles}.}  

To align the empirical specification with our theoretical framework, we consider the following partially linear panel model with interactive fixed effects:
\begin{align}
{R}_{it} &= \theta_0 {V}_{it-1} + l_0(\boldsymbol{X}_{it-1}) + \Gamma_i^{\prime} f_t + \lambda_i' f_t  +{U}_{it}, \label{eqn:R}\\
{V}_{it} &= {D}_{it} - m_0(\boldsymbol{X}_{it}) - \Gamma_i' f_t  - \phi_i^{\prime}f_t  \label{eqn:D}
\end{align}
where outcome variable $R_{it}$ denotes monthly excess returns of firm $i$ at time $t$, $D_{it-1}$ is treatment variable namely the lagged firm characteristic of interest (among bid-ask spread, change in inventory, dividend omission, size industry-adjusted, return on equity, scaled earnings forecast, turnover volatility), and $\boldsymbol{X}_{it}$ is a high-dimensional vector of  $p=89$  lagged firm characteristics, excluding the treatment variable.\footnote{These characteristics form a subset of the 102 variables compiled by \citet{green2017characteristics} for 1980-2014 and subsequently extended to 2022. Following \citet{rucker2025}, we exclude two characteristics that are constant over the sample period, nine variables constructed from past returns, and one characteristic that is zero for the majority of firms.} The structural parameter of interest $\theta_0$ measures the partial effect of the orthogonalised characteristic $D_{it}$ on monthly excess returns, after controlling for high-dimensional confounders and latent factor structures in a flexible manner with machine learning algorithms. The nuisance functions $l$ and $m$ accommodate potentially unknown function of firm characteristics on returns, consistent with the growing machine learning literature in asset pricing \citep[see][]{gu2020empirical}.

We estimate the Equations~\eqref{eqn:R}-\eqref{eqn:D} using conventional OLS estimator, the IFE estimator by \citet{bai2009panel}, the DML-FE estimator with the correlated random effects (CRE) approach by \cite{clarke2025double}, and our proposed panel DML-IFE estimator.%
\footnote{The empirical analysis is conducted in the statistical software \textsc{R} (version 4.4.0). IFE regressions are estimated using the \texttt{xtife} package \citep{xtife}, DML-FE using \texttt{xtdml} \citep{xtdml}, and DML-IFE using our \texttt{xtifedml} package \citep{xtifedml}, available on GitHub at \url{https://github.com/POLSEAN/xtifedml}.}
The DML estimations employ ML algorithms from three distinct families of learners (i.e., Lasso for penalised regression, gradient boosting for tree-based methods, and neural networks for deep learning) to predict the nuisance functions of the covariates.%
\footnote{For gradient boosting and neural networks, the covariate set consists solely of raw firm characteristics, as both learners are designed to capture nonlinearities automatically without requiring explicit interactions or polynomial expansions. In the case of neural networks, inputs are additionally standardised using min-max normalisation, following standard practice. Panel DML regressions with Lasso use an extended dictionary of nonlinear terms of the raw firm characteristics (i.e., polynomials up to order three and interaction terms of all covariates) so that the weak sparsity assumption holds. DML-FE regressions with Lasso and NNet, estimated using the CRE approach, also include the individual means of the included covariates as controls following \citet{clarke2025double}.} 
Our final sample consists of $N=29$ large-cap firms that are current or recent constituents of the Dow Jones Industrial Average, observed over $T=72$ monthly periods from April 2016 to March 2022.\footnote{Relative to \citet{rucker2025}, who use $T=60$, we extend the sample to exploit the full available time dimension.}  The data are drawn from CRSP, Compustat, and I/B/E/S. Detailed variable definitions and data construction procedures are provided in the Appendix.

Table~\ref{tab:main} reports estimates of the effect of various firm characteristics on monthly excess stock returns, across these four estimators. Overall, the results highlight the importance of both flexible nonlinear nuisance function estimation and the interactive fixed effects structure for reliable causal inference in asset pricing panels. 

\begin{table}[t!]
\centering
\caption{The Effect of Firm Characteristics on Stock Returns}\label{tab:main}
\scalebox{.7}{
\begin{threeparttable}	
\begin{tabular}{lcccccccc}
\vspace{-3mm}\\
\hline\hline
\vspace{-3mm}\\
Estimator: & OLS & IFE &\multicolumn{3}{c}{DML-FE}  & \multicolumn{3}{c}{DML-IFE} \\
\cmidrule(l{.35cm}r{.25cm}){4-6}\cmidrule(l{.35cm}r{.25cm}){7-9}
Learner:  & & & Lasso & Boosting & NNet & Lasso & Boosting &  NNet \\
\midrule
\multicolumn{9}{l}{\emph{Panel A: Bid-Ask Spread}} \\
 & 0.946*** & -0.207 & 0.161*** & 0.941** & 0.04 & -0.007 & -0.159 & -0.006 \\
 & (0.222) & (0.343) & (0.042) & (0.46) & (0.03) & (0.084) & (0.28) & (0.344) \\
\\[-0.5em]
Model RMSE &  &  & 0.564 & 0.602 & 0.590 & 0.542 & 0.569 & 0.570 \\
RMSE$_l$ &  &  & 0.070 & 0.074 & 0.074 & 0.069 & 0.070 & 0.069 \\
RMSE$_m$ &  &  & 0.024 & 0.008 & 0.080 & 0.014 & 0.006 & 0.008 \\
\\[-0.5em]

\midrule
\multicolumn{9}{l}{\emph{Panel B: Change in inventory}} \\
 & -0.144 & -0.07 & -0.107 & -0.032 & 0.041 & -0.037*** & -0.387* & -0.195** \\
 & (0.299) & (0.159) & (0.081) & (0.084) & (0.095) & (0.013) & (0.234) & (0.097) \\
\\[-0.5em]
Model RMSE &  &  & 0.564 & 0.599 & 0.597 & 0.538 & 0.571 & 0.569 \\
RMSE$_l$ &  &  & 0.069 & 0.074 & 0.074 & 0.069 & 0.070 & 0.069 \\
RMSE$_m$ &  &  & 0.035 & 0.036 & 0.054 & 0.126 & 0.018 & 0.023 \\
\\[-0.5em]

\midrule
\multicolumn{9}{l}{\emph{Panel C: Dividend omission}} \\
 & -0.057 & -0.041* & 0.000& -0.024 & -0.047*** & -0.024*** & -0.038** & -0.021 \\
 & (0.089) & (0.023) & (0.003) & (0.015) & (0.018) & (0.006) & (0.017) & (0.02) \\
\\[-0.5em]
Model RMSE &  &  & 0.564 & 0.628 & 0.593 & 0.542 & 0.620 & 0.571 \\
RMSE$_l$ &  &  & 0.069 & 0.075 & 0.073 & 0.069 & 0.074 & 0.069 \\
RMSE$_m$ &  &  & 0.477 & 0.167 & 0.093 & 0.104 & 0.143 & 0.087 \\
\\[-0.5em]

\midrule
\multicolumn{9}{l}{\emph{Panel D: Size (industry-adjusted)}} \\
 & 0.000& 0.002 & 0.000& 0.000& 0.004*** &  0.000*** & -0.001 & -0.002*** \\
 & (0.001) & (0.001) & (0.001) & (0.001) & (0.001) & (0.000) & (0.001) & (0.001) \\
\\[-0.5em]
Model RMSE &  &  & 0.565 & 0.610 & 0.606 & 0.539 & 0.598 & 0.563 \\
RMSE$_l$ &  &  & 0.069 & 0.073 & 0.074 & 0.069 & 0.071 & 0.069 \\
RMSE$_m$ &  &  & 4.154 & 2.343 & 4.468 & 24.706 & 1.699 & 2.776 \\
\\[-0.5em]

\midrule
\multicolumn{9}{l}{\emph{Panel E: Return on equity}} \\
& 0.016 & 0.012* & 0.000& 0.003 & -0.014 &  0.000*** & 0.014** & 0.008 \\
 & (0.016) & (0.007) & (0.006) & (0.004) & (0.01) & (0.000) & (0.006) & (0.008) \\
\\[-0.5em]
Model RMSE &  &  & 0.564 & 0.598 & 0.595 & 0.539 & 0.573 & 0.571 \\
RMSE$_l$ &  &  & 0.069 & 0.072 & 0.073 & 0.069 & 0.071 & 0.069 \\
RMSE$_m$ &  &  & 0.463 & 0.298 & 0.447 & 8.179 & 0.332 & 0.364 \\
\\[-0.5em]

\midrule
\multicolumn{9}{l}{\emph{Panel F: Scaled earnings forecast}} \\
& -0.099 & -0.004 & -0.067 & -0.193*** & -0.144*** & -0.092*** & -0.216*** & 0.003 \\
 & (0.114) & (0.081) & (0.061) & (0.023) & (0.04) & (0.013) & (0.074) & (0.077) \\
\\[-0.5em]
Model RMSE &  &  & 0.565 & 0.648 & 0.593 & 0.541 & 0.573 & 0.569 \\
RMSE$_l$ &  &  & 0.069 & 0.075 & 0.072 & 0.069 & 0.070 & 0.069 \\
RMSE$_m$ &  &  & 0.058 & 0.049 & 0.059 & 0.074 & 0.028 & 0.031 \\
\\[-0.5em]

\midrule
\multicolumn{9}{l}{\emph{Panel G: Turnover volatility}} \\
 & -0.005*** & -0.003* & -0.006** & -0.007*** & -0.004 & -0.005*** & -0.005*** & -0.001 \\
 & (0.001) & (0.002) & (0.003) & (0.002) & (0.003) & (0.001) & (0.001) & (0.002) \\
\\[-0.5em]
Model RMSE &  &  & 0.564 & 0.594 & 0.577 & 0.539 & 0.586 & 0.570 \\
RMSE$_l$ &  &  & 0.069 & 0.074 & 0.073 & 0.069 & 0.071 & 0.069 \\
RMSE$_m$ &  &  & 0.791 & 1.263 & 1.460 & 1.187 & 1.374 & 1.579 \\
\\[-0.5em]
\hline
\vspace{-3mm}\\
No. of firms & 29& 29 & 29 & 29 & 29 & 29 & 29 & 29 \\
No. of periods& 72& 72 & 72 & 72 & 72 & 72 & 72 & 72 \\
No. of controls& 89 & 89 & 89 & 89 & 89 & 89 & 89 & 89 \\
\midrule
\hline
\end{tabular}
\footnotesize
\textbf{Note:} The dependent variable in all panels is monthly excess stock returns; the treatment variables differ by panel.  The OLS column reports ordinarily least squares estimates; the IFE column the interactive fixed effects estimates following \cite{bai2009panel}; the DML-FE columns the DML estimates with fixed effects estimated using the correlated random effects (CRE) approach, as in \cite{clarke2025double}; the DML-IFE columns report DML estimates with interactive fixed effects as proposed in this paper. The DML regressions use different learners for predicting the nuisance functions. The set of control variables is a high-dimensional vector of lagged firm characteristics, excluding the treatment. Lasso uses an extended dictionary with polynomials and interactions. The DML estimation uses threefold block cross-fitting and the partialling-out score. The hyperparameters of the base learners are tuned with 100 random-search draws. Cluster-robust standard errors at the firm level are shown in parentheses. Significance levels: $^{*}p < 0.10$, $^{**}p < 0.05$, $^{***}p < 0.01$.
\end{threeparttable}
}
\end{table}

In Panel~A (bid-ask spread) of Table~\ref{tab:main}, the OLS estimator yields a large positive and strongly significant coefficient, consistent with the liquidity premium hypothesis of \citet{amihud1986asset}; investors demand higher returns as compensation for holding illiquid assets. However, this estimate is likely upward biased due to omitted unobserved interactive fixed effects, as supported by the statistically insignificant IFE estimates. Once nuisance functions are estimated flexibly via DML and the interactive factor structure is accounted for, the DML-IFE estimator under all learners produces small (negative) and statistically insignificant coefficients. This finding suggests that the apparent liquidity premium detected by OLS is attributable to unmodelled interactive fixed effects rather than a genuine causal relationship between bid-ask spreads and excess returns in this sample. The agreement between IFE and DML-IFE specifications across learners reinforces this conclusion.

In Panel~B (change in inventory), almost all specifications report a negative effect of inventory changes on excess returns. This is consistent with the investment-based asset pricing literature, which observed that firms accumulating excess inventories tend to be overvalued or to face deteriorating fundamentals, leading to lower subsequent returns \citep{thomas2002inventory}. OLS and IFE both produce negative but statistically insignificant coefficients, as does DML-FE under all learners. By contrast, our panel DML-IFE estimator yields statistically significant negative coefficients, which differ in magnitudes across learners. The relatively higher $RMSE_m$ under DML-IFE Lasso suggests some difficulty in partialling-out the treatment, possibly reflecting mild nonlinearity in how inventory changes relate to the control variables in a manner which is undetected by the provided Lasso dictionary. Nonetheless, the emergence of a significant effect exclusively under DML-IFE illustrates the advantage of our framework: the combination of flexible nuisance function estimation and interactive fixed effects elimination enables identification of a causal effect that would otherwise be masked by confounding.

In Panel~C (dividend omission), there is almost a general consensus among estimators of a negative causal effect of dividend omission on excess returns, which is theoretically consistent with signalling theories of dividend policy \citep{bhattacharya1979imperfect}. In practice, omitting dividends conveys adverse information about future firm prospects, depressing returns. The IFE estimator produces a small negative and borderline-significant coefficient, and this finding is broadly supported across DML specifications (i.e., DML-FE NNet and most DML-IFE learners). The relative consistency across estimators reduces concerns about unobserved interactive factor contamination and undetected nonlinearities by the IFE estimator. 

In Panel~D (size, industry-adjusted), the relationship between industry-adjusted firm size and excess returns is theoretically ambiguous, with the size effect documented in \citet{fama1993common} predicting a negative relationship between firm size and average returns, interpreted as compensation for a common size-related risk factor. OLS and IFE both yield statistically insignificant estimates. In contrast, among the DML specifications,  DML-IFE Lasso produces a statistically significant, but close-to-zero, coefficient while DML-IFE NNet yields a negative significant coefficient of opposite sign. When two flexible learners disagree on the direction of the estimated effect, this typically indicates either undetected nonlinearities in the relationship between size and returns or severe sensitivity of the partialling-out stage to how high-dimensional controls interact with the treatment. The extremely high $RMSE_m$ under DML-IFE Lasso confirms severe misfit in partialling-out the treatment, making this Lasso estimate unreliable. 

In Panel~E (return on equity), the positive causal effect of return on equity on excess returns found by most estimators is consistent with profitability-based asset pricing theories \citep{fama2015five}, where more profitable firms command higher expected returns as compensation for systematic risk. OLS yields a positive but statistically insignificant estimate, while IFE produces only a marginally significant result (at 10\% level). Across DML specifications, estimates remain close to zero, but being significant only in DML-IFE under Lasso and boosting. The large $RMSE_m$ value of DML-IFE Lasso indicate that the treatment cannot easily be partialled-out by the control variables, raising serious concerns about the reliability of the result. The fairly-significant IFE coefficient is likely to reflect a weak interactive factor loading rather than a genuine profitability premium, and the absence of a robust signal across DML specifications suggests that any apparent effect is fragile and sensitive to the treatment of unobserved heterogeneity.

In Panel~F (scaled earnings forecast), most of the estimates display a negative causal effect of scaled earnings forecasts on excess returns, which may reflect an overreaction mechanism. In the literature, optimistic analyst forecasts are usually associated with inflated current prices and subsequently lower realised returns \citep{bordalo2024belief}. OLS yields a negative but statistically insignificant estimate, and IFE likewise finds no significant effect. By contrast, DML specifications -- particularly those employing gradient boosting -- consistently recover a negative and statistically significant coefficient across both DML-FE and DML-IFE. This pattern suggests that once nonlinear relationships between controls and outcomes are flexibly modelled, a suppressed true effect emerges. The discrepancy between boosting and neural network under DML-IFE further implies that the control-treatment relationship exhibits nonlinear structure that neural network either overfits or partially absorbs into the nuisance estimation, pointing to residual functional form sensitivity in this panel.

In Panel~G (turnover volatility), there is a general agreement in a negative causal effect of turnover volatility on excess returns, which is consistent with investor attention and noise trading theories \citep{barber2008all}. For instance, high trading activity volatility is associated with overpricing and subsequent return reversals. The negative estimated coefficients are stable in sign and statistically significant across  all estimators and learners, including OLS, IFE, and both DML-FE and DML-IFE specifications. The uniformly low values of $RMSE_l$ and $RMSE_m$ confirm that the nuisance functions are well-predicted across all families of learners, reducing concern about functional form misspecification or undetected nonlinearities. This robustness across estimators and learners shows that, in this specific case, OLS delivers a reliable answer despite its stronger parametric assumptions, suggesting that the true underlying process may be linear.

Taken together, these results reveal three recurring patterns linked to the estimators considered. First, OLS estimates are systematically unreliable in this setting: they either overstate effects, as in Panel~A, where the apparent liquidity premium vanishes once interactive fixed effects are accounted for,  or fail to detect genuine effects, as in Panels~B and~F,where flexible nuisance estimation recovers significant coefficients that OLS misses entirely. This suggests that both omitted interactive factor structure and misspecification of the functional form of the control variables have substantial empirical consequences, and that relying on OLS alone would lead to substantially misleading inference. Second, the IFE estimator of \citet{bai2009panel} improves on OLS by accounting for unobserved interactive fixed effects, but it relies on the assumption of linearity in the control variables making it vulnerable to functional form misspecification, as most clearly illustrated in Panels~A,~B and~F. Third, the DML-IFE estimator, which combines flexible nonparametric nuisance estimation with interactive fixed effects elimination, consistently delivers estimates that are robust to both sources of confounding simultaneously. The advantage of this framework is most evident in Panels~B,~C and~F, where the combination of flexible learners and interactive fixed effects correction uncovers causal effects that all alternative estimators fail to detect. One exception to this overall picture concerns Panel~D, where conflicting signs across DML-IFE learners and large $RMSE_m$ values indicate that functional form fragility can persist even within the DML framework when the treatment variable is poorly identified after partialling-out a high-dimensional control set. Finally, Panel~G provides a useful benchmark: the robustness of the turnover volatility effect across all estimators and learners confirms that when a true causal relationship is sufficiently strong and easily identifiable, the choice of estimator matters little.

\section{Conclusion}\label{sec:conclusion}
This paper developed a doubly robust framework for causal inference in panel data settings with interactive fixed effects and high-dimensional covariates. By combining projection-based factor removal with Neyman-orthogonal score construction and cross-fitting, the proposed panel DML-IFE estimator achieves valid inference under latent common shocks without requiring parametric restrictions on the nuisance functions. Monte Carlo evidence confirmed the method's desirable finite-sample properties relative to conventional IFE estimator. The empirical application studying the effects of several treatment variables on monthly excess stock returns further demonstrated the method's practical relevance, remarking the importance of both flexible nonlinear nuisance estimation and the interactive fixed effects structure for reliable causal inference in asset pricing panels. 
As panel datasets grow in both cross-sectional and temporal dimensions and machine learning methods are becoming popular in applied work, we expect this framework to offer a practical and theoretically grounded toolkit for researchers seeking to combine flexible covariate modelling with rigorous treatment of unobserved heterogeneity.

Our simulation results demonstrate the importance of following an ensemble approach, as is commonly done in computer science because it is understood that no single learner will be superior across all applications.  Rather than relying on one learner, ensemble learning involves selecting the one (or a weighted average of the best results) that performs best across a range of learners.  In our simulation study, neural networks were consistently outperformed by Lasso and Boosting, and its performance was objectively poor for small sample sizes and a large number of irrelevant covariates. However, by using root mean square errors to choose the best-performing learner, analysts will avoid presenting biased point and interval estimates. 


Several promising directions remain for future research. The current framework assumes a fixed and known number of factors, and extending the method to data-driven factor number selection would enhance its applicability in settings where the latent structure remains uncertain. Additionally, generalising the approach to accommodate heterogeneous treatment effects, dynamic panels, or non-stationary factor loadings would broaden its scope and further align it with the demands of modern empirical work in economics and finance.

\bibliography{ref}
\bibliographystyle{apalike} 

\appendix
\newpage
\section{Asymptotics}

\subsection{Neyman Orthogonal Score Functions}\label{sec:score}

We derive the Neyman orthogonal score for the panel DML-IFE model via two
complementary approaches, both operating directly on the projected model
after applying the oracle projection $\Pi$.
Since $\Pi\mathbf{F}=\mathbf{0}$ annihilates the interactive fixed effect exactly, the factor loadings
$\{\boldsymbol{\lambda}_i,\boldsymbol{\phi}_i,\Gamma_i\}$ and common factors
$\{\mathbf{f}_t\}$ do not appear in the score derivation.
The unit of analysis is $W_i = (\mathbf{Y}_i,\mathbf{D}_i,\mathbf{X}_i)$
with $\mathbf{Y}_i,\mathbf{D}_i\in\mathbb{R}^T$ and $\mathbf{X}_i\in\mathbb{R}^{T\times p}$.
The nuisance parameters are $\eta = \bigl(l(\cdot),\,m(\cdot)\bigr)$
with true values $\eta_0 = (l_0,m_0)$ or, for the IV representation of the model (Remark 2.1), $\eta_0 = (g_0,m_0)$.

\subsubsection{Approach 1: Concentrate-out}\label{sec:score_co}

For the IV representation of the partially linear model, we follow \citet[][Lemma~2.5]{chernozhukov2018double} by defining the
unit-level criterion on the projected model:
\begin{equation}\label{eq:criterion}
  \ell(W_i;\,\theta,g)
  = -\tfrac{1}{2}\bigl\|\Pi_0\boldsymbol{Y}_i - \Pi_0\boldsymbol{D}_i\theta
    - \Pi_0\boldsymbol{g}(\mathbf{X}_i)\bigr\|^2
\end{equation}
The factor term $\mathbf{F}\boldsymbol{\lambda}_i$ does not appear because
$\Pi\mathbf{F}=\mathbf{0}$ exactly.
Let $g_{Y,\theta}$ concentrate out the nuisance by maximising
$\E[\ell(W_i;\theta,g)]$ over~$g$ for each~$\theta$.
The first-order condition requires
$\E\bigl[\Pi_0(\boldsymbol{Y}_i-\boldsymbol{D}_i\theta-\boldsymbol{g}_{Y,\theta}(\mathbf{X}_i))
         \mid\mathbf{X}_i\bigr]=\mathbf{0}_T$.
Substituting the model $\boldsymbol{Y}_i-\boldsymbol{D}_i\theta_0
= \boldsymbol{g}_0(\mathbf{X}_i)+\mathbf{F}\boldsymbol{\lambda}_i+\boldsymbol{U}_i$
and using $\Pi\mathbf{F}=\mathbf{0}$,
$\E[\Pi_0\boldsymbol U_i\mid\mathbf X_i]
=\E[\E[\Pi_0\boldsymbol U_i\mid\mathbf X_i,\boldsymbol D_i]\mid\mathbf X_i]
=\mathbf 0_T$ :
\begin{equation}
  \Pi_0\boldsymbol{g}_{Y,\theta_0}(\mathbf{X}_i) = \Pi_0\boldsymbol{g}_0(\mathbf{X}_i),
  \qquad
  \Pi_0\,\frac{d\,g_{Y,\theta}}{d\theta}\bigg|_{\theta_0}
  = -\Pi_0\boldsymbol{m}_0(\mathbf{X}_i).
\end{equation}
By Lemma~2.5 of \citet{chernozhukov2018double}, the Neyman orthogonal
score is $\psi_i = (d/d\theta)\,\ell(W_i;\theta,g_{Y,\theta})|_{\theta_0}$.
At $\theta_0$ the projected residual is
$\Pi_0\boldsymbol{Y}_i-\Pi_0\boldsymbol{D}_i\theta_0-\Pi_0\boldsymbol{g}_0
= \Pi_0\boldsymbol{U}_i$, and using
$\Pi_0(\boldsymbol{D}_i-\boldsymbol{m}_0(\mathbf{X}_i))=\Pi_0\boldsymbol{V}_i$:
\begin{equation}\label{eq:score_A1}
  \psi_i
  = (\Pi_0\boldsymbol{U}_i)'\bigl(-\Pi_0\boldsymbol{D}_i
    + \Pi_0\boldsymbol{m}_0(\mathbf{X}_i)\bigr)
  = -(\Pi_0\boldsymbol{U}_i)'\Pi_0\boldsymbol{V}_i
  = -\sum_{t=1}^T\widetilde{U}_{it}\,\widetilde{V}_{it},
\end{equation}
where $\widetilde{U}_{it}:=[\Pi_0\boldsymbol{U}_i]_t$ and
$\widetilde{V}_{it}:=[\Pi_0\boldsymbol{V}_i]_t=
[\Pi_0(\boldsymbol{D}_i-\boldsymbol{m}_0(\mathbf{X}_i))]_t$
are the oracle projected residuals.

\subsubsection{Approach 2: Projected Conditional Moment Restrictions}\label{sec:score_cmr}

For the partialled out representation, we follow Section~2.2.4 of \citet{chernozhukov2018double} for the projected
moment condition at the unit level
\begin{equation}\label{eq:moment_condition}
  \E\bigl[\Pi_0\{\boldsymbol{Y}_i - (\boldsymbol{D}_i-\vector{m}_0(\mathbf{X}_i))\theta_0
           - \boldsymbol{l}_0(\mathbf{X}_i)\}\mid\mathbf{X}_i,\boldsymbol{D}_i\bigr]
  = \mathbf{0}_T.
\end{equation}
Set $W_i=(\mathbf{Y}_i,\mathbf{D}_i,\mathbf{X}_i)$,
$R_i=(\mathbf{D}_i,\mathbf{X}_i)$, $Z_i=\mathbf{X}_i$, and let
$\gamma(W_i;\theta,l) = \Pi_0(\boldsymbol{Y}_i-\boldsymbol{D}_i\theta
-\boldsymbol{l}(\mathbf{X}_i))$ (a $T$-vector).
The Neyman orthogonal score takes the form
$\psi_i = \mu_0(R_i)\,\gamma(W_i;\theta_0,l_0)$,
where $\mu_0(R_i) = A(R_i)'\Omega(R_i)^{-}
- G(Z_i)\,J_l(R_i)'\,\Omega(R_i)^{-}$
and $\Omega(R_i)^{-}$ denotes any generalized inverse
(satisfying $\Omega\Omega^{-}\Omega=\Omega$), needed because the projected
residual lives in the $(T-r)$-dimensional range of $\Pi$ and $\Omega(R_i)$ is
rank-deficient. 
\begin{align}
  A(R_i) &= \partial_\theta\,\E\bigl[\Pi_0\{\boldsymbol{Y}_i - (\boldsymbol{D}_i-\vector{m}_0(\mathbf{X}_i))\theta_0
           - \boldsymbol{l}_0(\mathbf{X}_i)\}\mid R_i\bigr]
             \big|_{\theta=\theta_0}
         \notag\\
          &= -\Pi_0(\boldsymbol{D}_i-\vector{m}_0(\mathbf{X}_i)),
          \label{eq:A_cmr}\\[3pt]
    \Omega(R_i) &= \E\bigl[\Pi_0\boldsymbol{U}_i(\Pi_0\boldsymbol{U}_i)'
                \mid R_i\bigr]
             = \Pi_0\,\E[\boldsymbol U_i\boldsymbol U_i'\mid R_i]\,\Pi_0 \notag\\
            &= \sigma_0^2\,\Pi_0
           \quad\text{(under }\E[\boldsymbol U_i\boldsymbol U_i'\mid R_i]
             =\sigma_0^2 I_T\text{),}
             \label{eq:Omega_cmr}\\[3pt]
  J_l(R_i) &= \partial_{v1,v2}\,\E\bigl[\Pi_0\{\boldsymbol{Y}_i - v_1 - (\boldsymbol{D}_i-v_2)\theta_0
           \}\mid R_i\bigr]
                \big|_{(v_1,v_2)=(\vector{l}_0,\vector{m}_0)} \notag\\
             &= -\Pi_0\boldsymbol h, 
             \label{eq:Gamma_cmr}\\[3pt]
G(Z_i) &= \E\bigl[A(R_i)'\Omega^{-}J_l(R_i)\mid Z_i\bigr]
            \bigl(\E\bigl[J_l(R_i)'\Omega^{-}J_l(R_i)\mid Z_i\bigr]\bigr)^{-}
          \notag\\
          &=  \E\bigl[(-\Pi_0(\boldsymbol{D}_i- \vector{m}_0(\mathbf{X}_i)))'(\sigma_0^{-2}\Pi_0)(-\Pi_0\boldsymbol{h})\mid\mathbf X_i\bigr]  \E\bigl[(-\Pi_0\boldsymbol{h})'(\sigma_0^{-2}\Pi_0)(-\Pi_0\boldsymbol{h})\mid\mathbf X_i\bigr]^-\notag\\
          &=\sigma_0^{-2}\,\E\bigl[(\Pi_0(\boldsymbol{D}_i-\vector{m}_0(\mathbf{X}_i)))'\mid\mathbf X_i\bigr]\,\boldsymbol{h}\E\bigl[(-\Pi_0\boldsymbol{h})'(\sigma_0^{-2}\Pi_0)(-\Pi_0\boldsymbol{h})\mid\mathbf X_i\bigr]^-\notag\\
          &= \mathbf{0},
          \label{eq:G_cmr}
\end{align}
where $\vector{h}=(\boldsymbol{1}_T,-\boldsymbol{1}_T\theta_0)$.  Since
$\Pi_0\E[\boldsymbol{D}_i- \vector{m}_0(\mathbf{X}_i))\mid\mathbf X_i]
=(\Pi_0\mathbf F)\E[\boldsymbol\phi_i\mid\mathbf X_i]+\Pi_0\E[\boldsymbol V_i\mid\mathbf X_i]=\mathbf 0$
(using $\Pi_0\mathbf F=\mathbf 0$ and $\E[\boldsymbol V_i\mid\mathbf X_i]=\mathbf 0$).
Hence $G(Z_i)=\mathbf 0$. Using $A(R_i)'\Omega^{-}
= -(\Pi_0(\boldsymbol{D}_i-\vector{m}_0(\mathbf{X}_i)))'\sigma_0^{-2}\Pi_0 = -(\Pi_0(\boldsymbol{D}_i-\vector{m}_0(\mathbf{X}_i)))'$
(since $(\Pi_0(\boldsymbol{D}_i-\vector{m}_0(\mathbf{X}_i)))'\Pi_0=(\Pi_0(\boldsymbol{D}_i-\vector{m}_0(\mathbf{X}_i)))'$ and $\sigma_0^2=1$):
\begin{equation}\label{eq:mu_cmr}
  \mu_0(R_i)
  = A(R_i)'\Omega^{-}
  = -(\Pi_0(\boldsymbol{D}_i-\vector{m}_0(\mathbf{X}_i)))'.
\end{equation}

The Neyman orthogonal score is therefore
\begin{equation}\label{eq:score_A2}
  \psi_i
  = \mu_0(R_i)\,\gamma(W_i;\theta_0,l_0)
  = -(\Pi_0(\boldsymbol{D}_i-\vector{m}_0(\mathbf{X}_i)))'\Pi_0\boldsymbol{U}_i
  = -(\Pi_0\boldsymbol{V}_i)'\Pi_0\boldsymbol{U}_i
  = -\sum_{t=1}^T\widetilde{V}_{it}\,\widetilde{U}_{it},
\end{equation}

\begin{remark}[Sign convention and consistency of the two approaches]%
  \label{rem:sign_convention}
  \normalfont
  Both approaches yield $\psi_i = -\sum_t\widetilde{V}_{it}\widetilde{U}_{it}$,
  consistent up to an overall sign with the DML convention
  $\E[\psi_i]=0$, which is sign-invariant.
  Throughout the remainder of this appendix the positive sign convention is
  adopted: $\psi_{it} = \widetilde{V}_{it}\widetilde{U}_{it}$ where
  $\widetilde{V}_{it}=[\Pi_0(\boldsymbol{D}_i-\boldsymbol{m}_0(\mathbf{X}_i))]_t$
  and $\widetilde{U}_{it}=[\Pi_0(\boldsymbol{Y}_i-(\boldsymbol{D}_i-\vector{m}_0(\mathbf{X}_i))\theta_0
  -\boldsymbol{l}_0(\mathbf{X}_i))]_t$.
  In contrast to the unprojected model, where
  $G(Z)=\E[D_{it}\mid X_{it}]=m_0(X_{it})+\boldsymbol{\phi}_i'\mathbf{f}_t\neq 0$
  requires a non-trivial bias correction, the projected model achieves
  $G(Z_i)=\mathbf{0}_T$ exactly, confirming that the projection $\Pi$
  is precisely what makes the DML-IFE score Neyman orthogonal.
\end{remark}

\newpage
\section{Hyperparameter Tuning}\label{sec:tuning}
\renewcommand{\thetable}{B.\arabic{table}}
\setcounter{table}{0}


Hyperparameter tuning is critical for achieving accurate effect estimation in machine learning, regardless of the choice of base learners or estimators. 
The optimal configuration of hyperparameters of a base learner (obtained after \emph{hyperparameter tuning}) is essential to reach state-of-the-art performance in effect estimation, regardless the choice of estimators and learners \citep{machlanski2023}. By contrast, entirely relying on default hyperparameter values (e.g., suggested by statistical packages or the literature) severely compromises the ability of learners to reach their full potential, ultimately biasing the causal estimand \citep{bach2024hyper,machlanski2024}. 

The two most commonly adopted  strategies for hyper-parameter optimization are grid search and random search \citep{bergstra2012}. 
Hyperparameter optimization proceeds with trials of different configurations of values of the hyperparameters to tune. Resampling methods -- such as, cross-validation (CV) -- are used to evaluate the performance of the algorithm in terms of RMSE (when the hyperparameters are numeric). This procedure is repeated for several configurations until a stopping rule is applied (e.g., maximum number of evaluations). Finally, the configuration with the best performance (with, e.g.,  lowest RMSE) is selected and passed to the learner to train and test the model. 

The Monte Carlo simulations in Section~\ref{sec:mcsims} use hyperparameters tuned as shown in Table~\ref{tab:hyperpara}. 
\begin{table}[h!]
\centering
 \caption{Hyperparameter tuning} \vspace{.1cm}\label{tab:hyperpara}
\scalebox{.75}{
\begin{threeparttable}	
   \begin{tabular}{lllll}  
\hline\hline
\vspace{-4mm}\\
Learner&Hyperparamter& Value of parameter in set & Description\\	
\hline
\vspace{-2mm}\\
Lasso&\texttt{lambda.min}& cross-validated&$\lambda$ equivalent to minimum mean cross-validated error\\
\vspace{-2mm}\\



\vspace{-2mm}\\
Gradient Boosting&\texttt{lambda}&real value in \{0,2\}&L2 regularization term on weights.\\
&\texttt{maxdepth}&integer in \{2,10\}&Maximum depth of any node of the final tree.\\
&\texttt{nrounds}& 1000 &Number of decision trees in the final model\\

\vspace{-2mm}\\
NNET& \texttt{maxit}&500&Maximum number of iterations.\\
&\texttt{MaxNWts}&4000&The maximum allowable number of weights.\\
&\texttt{trace}&FALSE&Switch for tracing optimization..\\
&\texttt{size}& integer in \{2,5\}&Number of units in the hidden layer.\\
&\texttt{decay}&double in \{0,0.1\}& Parameter for weight decay.\\
\vspace{-2mm}\\
\hline
\end{tabular} 
\begin{tablenotes}[para,flushleft]	
\footnotesize{Note: The hyperparameters of the base learners chosen to model the nuisance functions are tuned in each Monte Carlo replication via 100 times of random search. 
}
\end{tablenotes}
\end{threeparttable}}
\end{table}

\newpage

\renewcommand{\theequation}{C.\arabic{equation}}
\setcounter{equation}{0}

\newpage
\clearpage
\section{Additional Tables}\label{sec:app_tabs}
\renewcommand{\thetable}{D.\arabic{table}}
\setcounter{table}{0}

\begin{table}[h!]
\centering
\caption{Monte Carlo Results: DGP 1, $N = 50$}
\label{tab:dgp1_n50}
\scalebox{.6}{
\begin{threeparttable}
\begin{tabular}{lcccccccccccc}
\vspace{-3mm}\\
\hline\hline
\vspace{-3mm}\\
 & \multicolumn{4}{c}{$T=30$} & \multicolumn{4}{c}{$T=50$} & \multicolumn{4}{c}{$T=100$} \\
\cmidrule(lr){2-5}\cmidrule(lr){6-9}\cmidrule(lr){10-13}
 & IFE & Lasso & Boosting & NNet &IFE & Lasso & Boosting &  NNet &IFE & Lasso & Boosting &  NNet\\
\hline
\multicolumn{13}{c}{\emph{Panel A: p = 50}}\\
Bias & 0.0022 & 0.0334 & -0.0190 & -0.1043 & 0.0030 & 0.0303 & -0.0077 & -0.0730 & 0.0000 & 0.0294 & 0.0045 & -0.0249 \\
Infeasible bias & 0.0002 & -0.0002 & -0.0498 & -0.0916 & -0.0011 & 0.0056 & -0.0364 & -0.0867 & -0.0010 & -0.0005 & -0.0257 & -0.0540 \\
MC RMSE & 0.0008 & 0.0020 & 0.0014 & 0.0397 & 0.0005 & 0.0012 & 0.0006 & 0.0395 & 0.0002 & 0.0011 & 0.0004 & 0.0179 \\
SD & 0.0276 & 0.0294 & 0.0318 & 0.1705 & 0.0218 & 0.0182 & 0.0225 & 0.1859 & 0.0143 & 0.0144 & 0.0198 & 0.1320 \\
SE & 0.0251 & 0.0269 & 0.0277 & 0.0409 & 0.0194 & 0.0207 & 0.0213 & 0.0306 & 0.4677 & 0.0147 & 0.0150 & 0.0212 \\
Model RMSE &   & 5.4295 & 6.0635 & 6.5533 &   & 7.0859 & 7.7317 & 8.2002 &   & 10.0970 & 10.6795 & 13.3505 \\
RMSE$_l$ &   & 1.4272 & 1.5177 & 1.9016 &   & 1.4377 & 1.5099 & 1.9006 &   & 1.4491 & 1.4918 & 1.7522 \\
RMSE$_m$ &   & 0.9936 & 1.0523 & 2.4180 &   & 1.0022 & 1.0490 & 1.2184 &   & 1.0101 & 1.0366 & 1.1136 \\
\multicolumn{13}{c}{\emph{Panel B: p = 100}}\\
Bias & 0.0054 & 0.0178 & -0.0469 & -0.2632 & 0.0054 & 0.0162 & -0.0278 & -0.1754 & 0.0015 & 0.0146 & -0.0138 & -0.1322 \\
Infeasible bias & 0.0016 & 0.0040 & -0.0638 & -0.2673 & -0.0019 & 0.0007 & -0.0431 & -0.2345 & 0.0005 & 0.0006 & -0.0279 & -0.1733 \\
MC RMSE & 0.0011 & 0.0010 & 0.0032 & 0.1125 & 0.0005 & 0.0007 & 0.0014 & 0.0721 & 0.0003 & 0.0005 & 0.0004 & 0.0652 \\
SD & 0.0325 & 0.0270 & 0.0313 & 0.2091 & 0.0225 & 0.0213 & 0.0247 & 0.2042 & 0.0160 & 0.0158 & 0.0139 & 0.2196 \\
SE & 0.0254 & 0.0264 & 0.0277 & 0.0460 & 0.0196 & 0.0203 & 0.0210 & 0.0480 & 0.5084 & 0.0144 & 0.0147 & 0.0305 \\
Model RMSE &   & 5.3846 & 6.1430 & 8.1503 &   & 6.9941 & 7.7701 & 16.4037 &   & 10.0250 & 10.6422 & 27.1115 \\
RMSE$_l$ &   & 1.4017 & 1.5045 & 2.5763 &   & 1.4144 & 1.4952 & 3.4322 &   & 1.4256 & 1.4705 & 2.7214 \\
RMSE$_m$ &   & 0.9848 & 1.0536 & 4.1320 &   & 0.9917 & 1.0443 & 4.3053 &   & 1.0014 & 1.0300 & 1.7294 \\
\multicolumn{13}{c}{\emph{Panel C: p = 300}}\\
Bias & 0.0269 & 0.0115 & -0.0811 & -0.6472 & 0.0105 & 0.0069 & -0.0533 & -0.5949 & 0.0020 & 0.0066 & -0.0298 & -0.3680 \\
Infeasible bias & 0.0027 & 0.0044 & -0.0821 & -0.6715 & -0.0022 & 0.0059 & -0.0597 & -0.5807 & -0.0009 & -0.0000 & -0.0357 & -0.3489 \\
MC RMSE & 0.0028 & 0.0008 & 0.0079 & 0.4777 & 0.0007 & 0.0004 & 0.0035 & 0.4397 & 0.0002 & 0.0002 & 0.0011 & 0.2021 \\
SD & 0.0453 & 0.0264 & 0.0366 & 0.2438 & 0.0234 & 0.0195 & 0.0261 & 0.2944 & 0.0146 & 0.0139 & 0.0149 & 0.2594 \\
SE & 0.0267 & 0.0262 & 0.0283 & 0.0473 & 0.0203 & 0.0204 & 0.0210 & 0.0447 & 0.5476 & 0.0141 & 0.0148 & 0.0227 \\
Model RMSE &   & 5.3505 & 6.2783 & 21.8995 &   & 6.9771 & 7.7711 & 22.0251 &   & 9.9585 & 10.7248 & 14.8021 \\
RMSE$_l$ &   & 1.3953 & 1.4981 & 4.2050 &   & 1.4038 & 1.4829 & 4.0315 &   & 1.4131 & 1.4656 & 1.7457 \\
RMSE$_m$ &   & 0.9842 & 1.0513 & 4.2337 &   & 0.9907 & 1.0428 & 8.4579 &   & 0.9977 & 1.0293 & 4.2476 \\
\hline
\end{tabular}
\begin{tablenotes}[para,flushleft]
\textbf{Note:} The figures in the table are average values and frequencies over 100 replications by estimation method, i.e.,  IFE by \citet{bai2009panel}, and our panel DML-IFE with Lasso, gradient boosting, and neural network. The true structural parameter $\theta$ is 1.  Standard errors in parenthesis are clustered at the firm level.  Additional details on panel DML-IFE estimation: cross-fitting with 5 folds. Lasso hyperparameter is selected from the model with minimum cross-validated error; gradient boosting and neural network hyperparameters are tuned with random search. 
\end{tablenotes}
\end{threeparttable}}
\end{table}


\begin{table}[t!]
\centering
\caption{Monte Carlo Results: DGP 2, $N = 50$}
\label{tab:dgp2_n50}
\scalebox{.6}{
\begin{threeparttable}
\begin{tabular}{lcccccccccccc}
\vspace{-3mm}\\
\hline\hline
\vspace{-3mm}\\
 & \multicolumn{4}{c}{$T=30$} & \multicolumn{4}{c}{$T=50$} & \multicolumn{4}{c}{$T=100$} \\
\cmidrule(lr){2-5}\cmidrule(lr){6-9}\cmidrule(lr){10-13}
 & IFE & Lasso & Boosting & NNet &IFE & Lasso & Boosting &  NNet &IFE & Lasso & Boosting &  NNet\\
\hline
\multicolumn{13}{c}{\emph{Panel A: p = 50}}\\
Bias & 0.0233 & 0.0845 & 0.0202 & 0.0918 & 0.0245 & 0.0763 & 0.0363 & 0.0007 & 0.0211 & 0.0714 & 0.0479 & 0.0481 \\
Infeasible bias & 0.1590 & 0.0195 & -0.0361 & 0.0293 & 0.1593 & 0.0201 & -0.0283 & -0.0643 & 0.1582 & 0.0089 & -0.0202 & -0.0161 \\
MC RMSE & 0.0013 & 0.0081 & 0.0017 & 0.9269 & 0.0010 & 0.0063 & 0.0019 & 0.0014 & 0.0006 & 0.0054 & 0.0027 & 0.0028 \\
SD & 0.0285 & 0.0314 & 0.0357 & 0.9632 & 0.0205 & 0.0215 & 0.0252 & 0.0373 & 0.0134 & 0.0168 & 0.0214 & 0.0221 \\
SE & 0.0243 & 0.0273 & 0.0284 & 0.2925 & 0.0188 & 0.0215 & 0.0219 & 0.0289 & 1.1560 & 0.0162 & 0.0167 & 0.0169 \\
Model RMSE &   & 5.6022 & 6.2653 & 29.3812 &   & 7.2927 & 7.8883 & 8.4005 &   & 10.3665 & 10.8566 & 11.2275 \\
RMSE$_l$ &   & 1.5196 & 1.5948 & 13.2142 &   & 1.5190 & 1.5796 & 1.8235 &   & 1.5243 & 1.5587 & 1.5769 \\
RMSE$_m$ &   & 1.0367 & 1.0930 & 5.1760 &   & 1.0384 & 1.0794 & 1.1012 &   & 1.0427 & 1.0656 & 1.0638 \\
\multicolumn{13}{c}{\emph{Panel B: p = 100}}\\
Bias & 0.0244 & 0.0553 & -0.0262 & -0.1729 & 0.0256 & 0.0465 & -0.0083 & -0.0175 & 0.0220 & 0.0424 & 0.0109 & 0.7428 \\
Infeasible bias & 0.1631 & 0.0246 & -0.0521 & -0.1657 & 0.1602 & 0.0147 & -0.0349 & -0.2949 & 0.1616 & 0.0103 & -0.0232 & -0.0844 \\
MC RMSE & 0.0014 & 0.0039 & 0.0022 & 0.3108 & 0.0011 & 0.0027 & 0.0007 & 2.4492 & 0.0007 & 0.0021 & 0.0004 & 53.6503 \\
SD & 0.0286 & 0.0292 & 0.0397 & 0.5326 & 0.0207 & 0.0225 & 0.0261 & 1.5728 & 0.0153 & 0.0173 & 0.0168 & 7.3236 \\
SE & 0.0245 & 0.0260 & 0.0283 & 0.2324 & 0.0190 & 0.0205 & 0.0210 & 0.2192 & 1.2388 & 0.0149 & 0.0151 & 0.6311 \\
Model RMSE &   & 5.5181 & 6.2626 & 33.9443 &   & 7.1475 & 7.8338 & 36.0573 &   & 10.2283 & 10.7261 & 55.3761 \\
RMSE$_l$ &   & 1.4695 & 1.5577 & 13.0126 &   & 1.4712 & 1.5299 & 15.5497 &   & 1.4780 & 1.5087 & 67.2524 \\
RMSE$_m$ &   & 1.0169 & 1.0838 & 2.8416 &   & 1.0184 & 1.0627 & 2.3772 &   & 1.0250 & 1.0479 & 1.1142 \\
\multicolumn{13}{c}{\emph{Panel C: p = 300}}\\
Bias & 0.0368 & 0.0377 & -0.0656 & -0.5997 & 0.0302 & 0.0283 & -0.0415 & -0.6332 & 0.0223 & 0.0218 & -0.0221 & -0.4913 \\
Infeasible bias & 0.1623 & 0.0222 & -0.0771 & -0.6764 & 0.1588 & 0.0181 & -0.0516 & -0.5780 & 0.1617 & 0.0089 & -0.0321 & -0.3565 \\
MC RMSE & 0.0029 & 0.0021 & 0.0053 & 0.4432 & 0.0014 & 0.0012 & 0.0025 & 0.4928 & 0.0007 & 0.0007 & 0.0007 & 0.3432 \\
SD & 0.0400 & 0.0253 & 0.0321 & 0.2907 & 0.0227 & 0.0211 & 0.0273 & 0.3046 & 0.0133 & 0.0138 & 0.0151 & 0.3206 \\
SE & 0.0262 & 0.0260 & 0.0283 & 0.0870 & 0.0196 & 0.0200 & 0.0211 & 0.0674 & 1.2986 & 0.0140 & 0.0145 & 0.0623 \\
Model RMSE &   & 5.4162 & 6.3371 & 26.8636 &   & 7.1201 & 7.8635 & 54.5157 &   & 10.1418 & 10.7538 & 50.9288 \\
RMSE$_l$ &   & 1.4448 & 1.5306 & 7.8902 &   & 1.4463 & 1.5069 & 7.7509 &   & 1.4476 & 1.4800 & 8.0969 \\
RMSE$_m$ &   & 1.0081 & 1.0707 & 4.2333 &   & 1.0109 & 1.0568 & 7.1533 &   & 1.0135 & 1.0388 & 8.3574 \\
\hline
\end{tabular}
\begin{tablenotes}[para,flushleft]
\textbf{Note:} The figures in the table are average values and frequencies over 100 replications by estimation method, i.e.,  IFE by \citet{bai2009panel}, and our panel DML-IFE with Lasso, gradient boosting, and neural network. The true structural parameter $\theta$ is 1.  Standard errors in parenthesis are clustered at the firm level.  Additional details on panel DML-IFE estimation: cross-fitting with 5 folds. Lasso hyperparameter is selected from the model with minimum cross-validated error; gradient boosting and neural network hyperparameters are tuned with random search. 
\end{tablenotes}
\end{threeparttable}}
\end{table}


\begin{table}[t!]
\centering
\caption{Monte Carlo Results: DGP 3, $N = 50$}
\label{tab:dgp3_n50}
\scalebox{.6}{
\begin{threeparttable}
\begin{tabular}{lcccccccccccc}
\vspace{-3mm}\\
\hline\hline
\vspace{-3mm}\\
 & \multicolumn{4}{c}{$T=30$} & \multicolumn{4}{c}{$T=50$} & \multicolumn{4}{c}{$T=100$} \\
\cmidrule(lr){2-5}\cmidrule(lr){6-9}\cmidrule(lr){10-13}
 & IFE & Lasso & Boosting & NNet &IFE & Lasso & Boosting &  NNet &IFE & Lasso & Boosting &  NNet\\
\hline
\multicolumn{13}{c}{\emph{Panel A: p = 50}}\\
Bias & 0.2025 & 0.1150 & 0.0320 & 0.0026 & 0.2022 & 0.1041 & 0.0705 & 0.0415 & 0.2007 & 0.0973 & 0.0897 & 0.0759 \\
Infeasible bias & 0.2049 & 0.0425 & -0.0197 & -0.0585 & 0.2028 & 0.0292 & -0.0131 & -0.0100 & 0.2028 & 0.0183 & -0.0160 & -0.0069 \\
MC RMSE & 0.0420 & 0.0143 & 0.0036 & 0.0123 & 0.0413 & 0.0115 & 0.0076 & 0.0033 & 0.0405 & 0.0098 & 0.0096 & 0.0064 \\
SD & 0.0315 & 0.0331 & 0.0514 & 0.1117 & 0.0215 & 0.0267 & 0.0516 & 0.0402 & 0.0149 & 0.0193 & 0.0399 & 0.0252 \\
SE & 0.0256 & 0.0278 & 0.0313 & 0.0581 & 0.0198 & 0.0220 & 0.0238 & 0.0246 & 0.6809 & 0.0172 & 0.0173 & 0.0179 \\
Model RMSE &   & 5.6520 & 7.1596 & 7.6504 &   & 7.3501 & 8.7670 & 8.5557 &   & 10.4225 & 11.7802 & 11.1507 \\
RMSE$_l$ &   & 1.5519 & 1.7595 & 2.4169 &   & 1.5532 & 1.7343 & 1.6867 &   & 1.5518 & 1.6931 & 1.6139 \\
RMSE$_m$ &   & 1.0371 & 1.1476 & 1.1998 &   & 1.0458 & 1.1314 & 1.1198 &   & 1.0472 & 1.1142 & 1.0768 \\
\multicolumn{13}{c}{\emph{Panel B: p = 100}}\\
Bias & 0.2055 & 0.0847 & 0.0080 & -0.1752 & 0.2051 & 0.0665 & -0.0020 & -0.1237 & 0.2015 & 0.0577 & 0.0169 & -0.0228 \\
Infeasible bias & 0.2066 & 0.0442 & -0.0167 & -0.2776 & 0.2044 & 0.0261 & -0.0292 & -0.1337 & 0.2051 & 0.0162 & -0.0286 & -0.0451 \\
MC RMSE & 0.0433 & 0.0080 & 0.0018 & 0.1101 & 0.0426 & 0.0050 & 0.0021 & 0.0649 & 0.0409 & 0.0036 & 0.0016 & 0.0022 \\
SD & 0.0327 & 0.0294 & 0.0424 & 0.2831 & 0.0237 & 0.0238 & 0.0457 & 0.2238 & 0.0171 & 0.0157 & 0.0365 & 0.0417 \\
SE & 0.0257 & 0.0266 & 0.0309 & 0.1194 & 0.0200 & 0.0207 & 0.0234 & 0.0621 & 0.7395 & 0.0151 & 0.0164 & 0.0184 \\
Model RMSE &   & 5.6052 & 7.1075 & 23.0092 &   & 7.2381 & 8.9872 & 15.6564 &   & 10.2245 & 12.1603 & 12.0028 \\
RMSE$_l$ &   & 1.5100 & 1.7397 & 5.8703 &   & 1.4975 & 1.6965 & 3.9467 &   & 1.4913 & 1.6518 & 1.6250 \\
RMSE$_m$ &   & 1.0237 & 1.1441 & 1.4443 &   & 1.0238 & 1.1250 & 1.5707 &   & 1.0237 & 1.1020 & 1.1093 \\
\multicolumn{13}{c}{\emph{Panel C: p = 300}}\\
Bias & 0.2302 & 0.0584 & 0.0078 & -0.5650 & 0.2096 & 0.0450 & -0.0206 & -0.5238 & 0.2001 & 0.0313 & -0.0338 & -0.3594 \\
Infeasible bias & 0.2093 & 0.0443 & 0.0035 & -0.5624 & 0.2038 & 0.0305 & -0.0305 & -0.5765 & 0.2043 & 0.0163 & -0.0432 & -0.4312 \\
MC RMSE & 0.0555 & 0.0041 & 0.0020 & 0.4572 & 0.0445 & 0.0026 & 0.0032 & 0.3732 & 0.0403 & 0.0012 & 0.0026 & 0.2716 \\
SD & 0.0502 & 0.0272 & 0.0439 & 0.3733 & 0.0248 & 0.0232 & 0.0526 & 0.3160 & 0.0153 & 0.0157 & 0.0383 & 0.3793 \\
SE & 0.0265 & 0.0259 & 0.0316 & 0.0787 & 0.0205 & 0.0201 & 0.0242 & 0.0894 & 0.7885 & 0.0143 & 0.0164 & 0.0996 \\
Model RMSE &   & 5.5090 & 7.2322 & 20.3714 &   & 7.1523 & 9.1561 & 42.5803 &   & 10.1259 & 12.3876 & 43.6771 \\
RMSE$_l$ &   & 1.4654 & 1.7500 & 6.2577 &   & 1.4601 & 1.6959 & 10.5439 &   & 1.4539 & 1.6300 & 9.6826 \\
RMSE$_m$ &   & 1.0070 & 1.1392 & 4.9242 &   & 1.0091 & 1.1200 & 4.0888 &   & 1.0117 & 1.1013 & 5.3613 \\
\hline
\end{tabular}
\begin{tablenotes}[para,flushleft]
\textbf{Note:} The figures in the table are average values and frequencies over 100 replications by estimation method, i.e.,  IFE by \citet{bai2009panel}, and our panel DML-IFE with Lasso, gradient boosting, and neural network. The true structural parameter $\theta$ is 1.  Standard errors in parenthesis are clustered at the firm level.  Additional details on panel DML-IFE estimation: cross-fitting with 5 folds. Lasso hyperparameter is selected from the model with minimum cross-validated error; gradient boosting and neural network hyperparameters are tuned with random search. 
\end{tablenotes}
\end{threeparttable}}
\end{table}


\begin{table}[t!]
\centering
\caption{Monte Carlo Results: DGP 1, $N = 100$}
\label{tab:dgp1_n100}
\scalebox{.6}{
\begin{threeparttable}
\begin{tabular}{lcccccccccccc}
\vspace{-3mm}\\
\hline\hline
\vspace{-3mm}\\
 & \multicolumn{4}{c}{$T=30$} & \multicolumn{4}{c}{$T=50$} & \multicolumn{4}{c}{$T=100$} \\
\cmidrule(lr){2-5}\cmidrule(lr){6-9}\cmidrule(lr){10-13}
 & IFE & Lasso & Boosting & NNet &IFE & Lasso & Boosting &  NNet &IFE & Lasso & Boosting &  NNet\\
\hline
\multicolumn{13}{c}{\emph{Panel A: p = 50}}\\
Bias & 0.0027 & 0.0298 & -0.0057 & -0.0495 & -0.0022 & 0.0317 & 0.0086 & -0.0295 & 0.0019 & 0.0305 & 0.0147 & -0.0150 \\
Infeasible bias & -0.0015 & 0.0032 & -0.0352 & -0.0809 & -0.0005 & -0.0006 & -0.0196 & -0.0378 & -0.0004 & 0.0007 & -0.0136 & -0.0355 \\
MC RMSE & 0.0005 & 0.0013 & 0.0006 & 0.0264 & 0.0003 & 0.0012 & 0.0004 & 0.0317 & 0.0001 & 0.0010 & 0.0003 & 0.0228 \\
SD & 0.0230 & 0.0200 & 0.0239 & 0.1556 & 0.0171 & 0.0155 & 0.0168 & 0.1766 & 0.0102 & 0.0104 & 0.0101 & 0.1512 \\
SE & 0.0180 & 0.0190 & 0.0196 & 0.0303 & 0.0139 & 0.0148 & 0.0149 & 0.0241 & 0.0098 & 0.0107 & 0.0107 & 0.0136 \\
Model RMSE &   & 5.4221 & 5.8417 & 7.7902 &   & 7.0903 & 7.4619 & 7.7548 &   & 10.0615 & 10.4682 & 10.6350 \\
RMSE$_l$ &   & 1.4217 & 1.4767 & 1.8769 &   & 1.4371 & 1.4792 & 1.8925 &   & 1.4492 & 1.4768 & 1.5247 \\
RMSE$_m$ &   & 0.9915 & 1.0266 & 1.0970 &   & 1.0012 & 1.0260 & 1.1495 &   & 1.0096 & 1.0276 & 1.0795 \\
\multicolumn{13}{c}{\emph{Panel B: p = 100}}\\
Bias & 0.0018 & 0.0155 & -0.0248 & -0.1703 & -0.0006 & 0.0147 & -0.0141 & -0.1556 & 0.0004 & 0.0162 & -0.0016 & -0.0674 \\
Infeasible bias & 0.0011 & 0.0023 & -0.0381 & -0.1826 & -0.0010 & 0.0031 & -0.0290 & -0.1452 & 0.0012 & 0.0017 & -0.0163 & -0.0687 \\
MC RMSE & 0.0004 & 0.0005 & 0.0011 & 0.0700 & 0.0003 & 0.0004 & 0.0005 & 0.0963 & 0.0001 & 0.0004 & 0.0001 & 0.0392 \\
SD & 0.0190 & 0.0175 & 0.0212 & 0.2035 & 0.0171 & 0.0144 & 0.0173 & 0.2699 & 0.0107 & 0.0103 & 0.0112 & 0.1871 \\
SE & 0.0182 & 0.0189 & 0.0197 & 0.0328 & 0.0140 & 0.0147 & 0.0148 & 0.0260 & 0.0098 & 0.0101 & 0.0105 & 0.0160 \\
Model RMSE &   & 5.3409 & 5.8429 & 6.6862 &   & 7.0454 & 7.4249 & 8.1710 &   & 9.9650 & 10.3828 & 11.4868 \\
RMSE$_l$ &   & 1.3952 & 1.4622 & 1.9487 &   & 1.4132 & 1.4581 & 2.3054 &   & 1.4262 & 1.4535 & 1.7644 \\
RMSE$_m$ &   & 0.9798 & 1.0220 & 1.4732 &   & 0.9910 & 1.0215 & 2.1995 &   & 1.0016 & 1.0186 & 1.2756 \\
\multicolumn{13}{c}{\emph{Panel C: p = 300}}\\
Bias & 0.0011 & 0.0072 & -0.0460 & -0.5464 & 0.0035 & 0.0060 & -0.0298 & -0.3642 & 0.0010 & 0.0051 & -0.0156 & -0.1814 \\
Infeasible bias & 0.0017 & 0.0059 & -0.0529 & -0.4610 & -0.0001 & 0.0037 & -0.0341 & -0.3635 & -0.0017 & 0.0021 & -0.0205 & -0.2002 \\
MC RMSE & 0.0004 & 0.0005 & 0.0026 & 0.3807 & 0.0003 & 0.0003 & 0.0012 & 0.1981 & 0.0001 & 0.0001 & 0.0004 & 0.0499 \\
SD & 0.0207 & 0.0215 & 0.0229 & 0.2882 & 0.0157 & 0.0151 & 0.0180 & 0.2571 & 0.0107 & 0.0105 & 0.0104 & 0.1309 \\
SE & 0.0188 & 0.0189 & 0.0198 & 0.0311 & 0.0143 & 0.0145 & 0.0150 & 0.0323 & 0.0100 & 0.0100 & 0.0103 & 0.0197 \\
Model RMSE &   & 5.3313 & 5.9230 & 10.2077 &   & 6.9639 & 7.5341 & 11.7699 &   & 9.9384 & 10.3785 & 12.6503 \\
RMSE$_l$ &   & 1.3866 & 1.4551 & 2.8263 &   & 1.3966 & 1.4497 & 2.7098 &   & 1.4089 & 1.4405 & 1.7097 \\
RMSE$_m$ &   & 0.9778 & 1.0242 & 6.7531 &   & 0.9855 & 1.0168 & 14.5767 &   & 0.9939 & 1.0130 & 1.3969 \\
\hline
\end{tabular}
\begin{tablenotes}[para,flushleft]
\textbf{Note:} The figures in the table are average values and frequencies over 100 replications by estimation method, i.e.,  IFE by \citet{bai2009panel}, and our panel DML-IFE with Lasso, gradient boosting, and neural network. The true structural parameter $\theta$ is 1.  Standard errors in parenthesis are clustered at the firm level.  Additional details on panel DML-IFE estimation: cross-fitting with 5 folds. Lasso hyperparameter is selected from the model with minimum cross-validated error; gradient boosting and neural network hyperparameters are tuned with random search. 
\end{tablenotes}
\end{threeparttable}}
\end{table}


\begin{table}[t!]
\centering
\caption{Monte Carlo Results: DGP 2, $N = 100$}
\label{tab:dgp2_n100}
\scalebox{.6}{
\begin{threeparttable}
\begin{tabular}{lcccccccccccc}
\vspace{-3mm}\\
\hline\hline
\vspace{-3mm}\\
 & \multicolumn{4}{c}{$T=30$} & \multicolumn{4}{c}{$T=50$} & \multicolumn{4}{c}{$T=100$} \\
\cmidrule(lr){2-5}\cmidrule(lr){6-9}\cmidrule(lr){10-13}
 & IFE & Lasso & Boosting & NNet &IFE & Lasso & Boosting &  NNet &IFE & Lasso & Boosting &  NNet\\
\hline
\multicolumn{13}{c}{\emph{Panel A: p = 50}}\\
Bias & 0.0243 & 0.0833 & 0.0375 & 0.0248 & 0.0197 & 0.0783 & 0.0518 & 0.0486 & 0.0235 & 0.0743 & 0.0555 & 0.0649 \\
Infeasible bias & 0.1599 & 0.0225 & -0.0212 & -0.0165 & 0.1594 & 0.0138 & -0.0109 & -0.0140 & 0.1599 & 0.0100 & -0.0086 & -0.0008 \\
MC RMSE & 0.0011 & 0.0076 & 0.0021 & 0.0018 & 0.0007 & 0.0065 & 0.0031 & 0.0030 & 0.0007 & 0.0057 & 0.0033 & 0.0045 \\
SD & 0.0217 & 0.0251 & 0.0266 & 0.0353 & 0.0165 & 0.0201 & 0.0196 & 0.0245 & 0.0104 & 0.0147 & 0.0138 & 0.0163 \\
SE & 0.0175 & 0.0193 & 0.0196 & 0.0215 & 0.0135 & 0.0156 & 0.0155 & 0.0163 & 0.0095 & 0.0118 & 0.0117 & 0.0123 \\
Model RMSE &   & 5.5923 & 5.9657 & 6.2887 &   & 7.2961 & 7.6135 & 7.8667 &   & 10.3446 & 10.6424 & 10.8221 \\
RMSE$_l$ &   & 1.5164 & 1.5512 & 1.6030 &   & 1.5199 & 1.5503 & 1.5693 &   & 1.5268 & 1.5384 & 1.5574 \\
RMSE$_m$ &   & 1.0353 & 1.0633 & 1.0831 &   & 1.0385 & 1.0586 & 1.0583 &   & 1.0433 & 1.0536 & 1.0497 \\
\multicolumn{13}{c}{\emph{Panel B: p = 100}}\\
Bias & 0.0231 & 0.0518 & 0.0032 & -0.2756 & 0.0207 & 0.0454 & 0.0081 & -0.0507 & 0.0219 & 0.0429 & 0.0226 & 0.0133 \\
Infeasible bias & 0.1599 & 0.0196 & -0.0258 & -0.4256 & 0.1599 & 0.0153 & -0.0203 & -0.0856 & 0.1610 & 0.0111 & -0.0107 & -0.0143 \\
MC RMSE & 0.0008 & 0.0030 & 0.0006 & 0.4866 & 0.0007 & 0.0023 & 0.0004 & 0.0829 & 0.0006 & 0.0020 & 0.0007 & 0.0007 \\
SD & 0.0177 & 0.0187 & 0.0247 & 0.6440 & 0.0161 & 0.0167 & 0.0177 & 0.2848 & 0.0102 & 0.0117 & 0.0123 & 0.0224 \\
SE & 0.0177 & 0.0187 & 0.0193 & 0.1789 & 0.0136 & 0.0148 & 0.0148 & 0.0692 & 0.0095 & 0.0105 & 0.0107 & 0.0114 \\
Model RMSE &   & 5.4690 & 5.9338 & 47.4509 &   & 7.2027 & 7.5263 & 32.9518 &   & 10.1772 & 10.4693 & 11.1499 \\
RMSE$_l$ &   & 1.4619 & 1.5103 & 10.1473 &   & 1.4704 & 1.4966 & 5.5668 &   & 1.4770 & 1.4881 & 1.5362 \\
RMSE$_m$ &   & 1.0120 & 1.0479 & 3.8245 &   & 1.0172 & 1.0423 & 1.7134 &   & 1.0239 & 1.0339 & 1.0435 \\
\multicolumn{13}{c}{\emph{Panel C: p = 300}}\\
Bias & 0.0220 & 0.0301 & -0.0251 & -0.5400 & 0.0244 & 0.0244 & -0.0163 & -0.4573 & 0.0225 & 0.0230 & -0.0041 & -0.3321 \\
Infeasible bias & 0.1611 & 0.0240 & -0.0373 & -0.5006 & 0.1604 & 0.0163 & -0.0259 & -0.4680 & 0.1597 & 0.0119 & -0.0160 & -0.3057 \\
MC RMSE & 0.0008 & 0.0014 & 0.0012 & 0.4114 & 0.0009 & 0.0009 & 0.0006 & 0.3947 & 0.0006 & 0.0007 & 0.0001 & 0.6836 \\
SD & 0.0190 & 0.0216 & 0.0231 & 0.3478 & 0.0162 & 0.0166 & 0.0173 & 0.4330 & 0.0104 & 0.0115 & 0.0111 & 0.7610 \\
SE & 0.0182 & 0.0186 & 0.0196 & 0.0685 & 0.0138 & 0.0143 & 0.0148 & 0.0821 & 0.0096 & 0.0100 & 0.0103 & 0.2018 \\
Model RMSE &   & 5.4482 & 5.9734 & 21.9787 &   & 7.0752 & 7.5860 & 43.2744 &   & 10.1067 & 10.4108 & 37.4828 \\
RMSE$_l$ &   & 1.4331 & 1.4913 & 7.9847 &   & 1.4359 & 1.4746 & 7.8062 &   & 1.4464 & 1.4570 & 26.2334 \\
RMSE$_m$ &   & 1.0031 & 1.0439 & 5.2040 &   & 1.0063 & 1.0317 & 6.1921 &   & 1.0109 & 1.0216 & 3.4140 \\
\hline
\end{tabular}
\begin{tablenotes}[para,flushleft]
\textbf{Note:} The figures in the table are average values and frequencies over 100 replications by estimation method, i.e.,  IFE by \citet{bai2009panel}, and our panel DML-IFE with Lasso, gradient boosting, and neural network. The true structural parameter $\theta$ is 1.  Standard errors in parenthesis are clustered at the firm level.  Additional details on panel DML-IFE estimation: cross-fitting with 5 folds. Lasso hyperparameter is selected from the model with minimum cross-validated error; gradient boosting and neural network hyperparameters are tuned with random search. 
\end{tablenotes}
\end{threeparttable}}
\end{table}


\begin{table}[t!]
\centering
\caption{Monte Carlo Results: DGP 3, $N = 100$}
\label{tab:dgp3_n100}
\scalebox{.6}{
\begin{threeparttable}
\begin{tabular}{lcccccccccccc}
\vspace{-3mm}\\
\hline\hline
\vspace{-3mm}\\
 & \multicolumn{4}{c}{$T=30$} & \multicolumn{4}{c}{$T=50$} & \multicolumn{4}{c}{$T=100$} \\
\cmidrule(lr){2-5}\cmidrule(lr){6-9}\cmidrule(lr){10-13}
 & IFE & Lasso & Boosting & NNet &IFE & Lasso & Boosting &  NNet &IFE & Lasso & Boosting &  NNet\\
\hline
\multicolumn{13}{c}{\emph{Panel A: p = 50}}\\
Bias & 0.2017 & 0.1185 & 0.1008 & 0.0715 & 0.1964 & 0.1021 & 0.1115 & 0.0830 & 0.2008 & 0.0940 & 0.0637 & 0.0863 \\
Infeasible bias & 0.2062 & 0.0436 & 0.0106 & 0.0103 & 0.2027 & 0.0270 & 0.0093 & 0.0066 & 0.2056 & 0.0142 & -0.0263 & -0.0121 \\
MC RMSE & 0.0412 & 0.0148 & 0.0121 & 0.0061 & 0.0389 & 0.0109 & 0.0132 & 0.0076 & 0.0404 & 0.0091 & 0.0048 & 0.0080 \\
SD & 0.0219 & 0.0274 & 0.0444 & 0.0314 & 0.0180 & 0.0219 & 0.0271 & 0.0276 & 0.0102 & 0.0171 & 0.0268 & 0.0241 \\
SE & 0.0187 & 0.0199 & 0.0212 & 0.0218 & 0.0144 & 0.0157 & 0.0162 & 0.0170 & 0.0102 & 0.0121 & 0.0123 & 0.0130 \\
Model RMSE &   & 5.6146 & 6.5249 & 6.2348 &   & 7.3195 & 8.1022 & 8.0434 &   & 10.3903 & 11.5806 & 11.0942 \\
RMSE$_l$ &   & 1.5505 & 1.7057 & 1.6582 &   & 1.5462 & 1.6821 & 1.6211 &   & 1.5463 & 1.6469 & 1.6057 \\
RMSE$_m$ &   & 1.0357 & 1.1113 & 1.0990 &   & 1.0414 & 1.1056 & 1.0724 &   & 1.0457 & 1.0982 & 1.0632 \\
\multicolumn{13}{c}{\emph{Panel B: p = 100}}\\
Bias & 0.2001 & 0.0842 & 0.0380 & -0.0545 & 0.1985 & 0.0699 & 0.0395 & -0.0139 & 0.1993 & 0.0578 & 0.0204 & 0.0307 \\
Infeasible bias & 0.2049 & 0.0442 & 0.0026 & -0.1211 & 0.2054 & 0.0286 & -0.0088 & -0.0408 & 0.2071 & 0.0163 & -0.0209 & -0.0061 \\
MC RMSE & 0.0405 & 0.0075 & 0.0032 & 0.0137 & 0.0397 & 0.0052 & 0.0030 & 0.0013 & 0.0399 & 0.0035 & 0.0008 & 0.0017 \\
SD & 0.0222 & 0.0206 & 0.0423 & 0.1042 & 0.0160 & 0.0164 & 0.0383 & 0.0331 & 0.0115 & 0.0106 & 0.0190 & 0.0272 \\
SE & 0.0186 & 0.0192 & 0.0213 & 0.0421 & 0.0145 & 0.0148 & 0.0162 & 0.0175 & 0.0102 & 0.0105 & 0.0114 & 0.0123 \\
Model RMSE &   & 5.5585 & 6.6860 & 8.0683 &   & 7.2072 & 8.4261 & 8.5862 &   & 10.2149 & 11.5032 & 11.3439 \\
RMSE$_l$ &   & 1.4991 & 1.6767 & 2.6253 &   & 1.4944 & 1.6476 & 1.6292 &   & 1.4887 & 1.5975 & 1.5653 \\
RMSE$_m$ &   & 1.0170 & 1.1036 & 1.1859 &   & 1.0202 & 1.0946 & 1.1052 &   & 1.0240 & 1.0842 & 1.0569 \\
\multicolumn{13}{c}{\emph{Panel C: p = 300}}\\
Bias & 0.2019 & 0.0579 & -0.0034 & -0.4585 & 0.2014 & 0.0411 & -0.0033 & -0.4774 & 0.2011 & 0.0299 & -0.0001 & -0.2957 \\
Infeasible bias & 0.2067 & 0.0438 & -0.0097 & -0.5254 & 0.2066 & 0.0266 & -0.0133 & -0.3900 & 0.2056 & 0.0156 & -0.0139 & -0.2942 \\
MC RMSE & 0.0413 & 0.0037 & 0.0020 & 0.3359 & 0.0409 & 0.0019 & 0.0015 & 0.3567 & 0.0406 & 0.0010 & 0.0003 & 0.1806 \\
SD & 0.0235 & 0.0185 & 0.0448 & 0.3562 & 0.0180 & 0.0148 & 0.0391 & 0.3608 & 0.0118 & 0.0124 & 0.0188 & 0.3068 \\
SE & 0.0192 & 0.0190 & 0.0218 & 0.0994 & 0.0147 & 0.0143 & 0.0164 & 0.0718 & 0.0103 & 0.0101 & 0.0109 & 0.0253 \\
Model RMSE &   & 5.5070 & 6.9029 & 50.8020 &   & 7.1360 & 8.6580 & 38.8874 &   & 10.0876 & 11.5628 & 31.3949 \\
RMSE$_l$ &   & 1.4604 & 1.6743 & 10.3876 &   & 1.4528 & 1.6357 & 8.2085 &   & 1.4495 & 1.5806 & 3.5884 \\
RMSE$_m$ &   & 1.0021 & 1.1083 & 5.7521 &   & 1.0055 & 1.0892 & 9.7162 &   & 1.0095 & 1.0770 & 6.1129 \\
\hline
\end{tabular}
\begin{tablenotes}[para,flushleft]
\textbf{Note:} The figures in the table are average values and frequencies over 100 replications by estimation method, i.e.,  IFE by \citet{bai2009panel}, and our panel DML-IFE with Lasso, gradient boosting, and neural network. The true structural parameter $\theta$ is 1.  Standard errors in parenthesis are clustered at the firm level.  Additional details on panel DML-IFE estimation: cross-fitting with 5 folds. Lasso hyperparameter is selected from the model with minimum cross-validated error; gradient boosting and neural network hyperparameters are tuned with random search. 
\end{tablenotes}
\end{threeparttable}}
\end{table}


\begin{table}[t!]
\centering
\caption{Monte Carlo Results: DGP 1, $N = 500$}
\label{tab:dgp1_n500}
\scalebox{.6}{
\begin{threeparttable}
\begin{tabular}{lcccccccccccc}
\vspace{-3mm}\\
\hline\hline
\vspace{-3mm}\\
 & \multicolumn{4}{c}{$T=30$} & \multicolumn{4}{c}{$T=50$} & \multicolumn{4}{c}{$T=100$} \\
\cmidrule(lr){2-5}\cmidrule(lr){6-9}\cmidrule(lr){10-13}
 & IFE & Lasso & Boosting & NNet &IFE & Lasso & Boosting &  NNet &IFE & Lasso & Boosting &  NNet\\
\hline
\multicolumn{13}{c}{\emph{Panel A: p = 50}}\\
Bias & -0.0005 & 0.0302 & 0.0189 & -0.0048 & 0.0001 & 0.0276 & 0.0227 & 0.0221 & -0.0004 & 0.0308 & 0.0253 & -0.0164 \\
Infeasible bias & -0.0003 & 0.0004 & -0.0095 & -0.0329 & -0.0001 & 0.0009 & -0.0060 & -0.0077 & -0.0003 & -0.0003 & -0.0031 & -0.0310 \\
MC RMSE & 0.0001 & 0.0010 & 0.0005 & 0.0130 & 0.0000 & 0.0008 & 0.0006 & 0.0006 & 0.0000 & 0.0010 & 0.0007 & 0.0193 \\
SD & 0.0086 & 0.0110 & 0.0105 & 0.1145 & 0.0063 & 0.0085 & 0.0093 & 0.0106 & 0.0054 & 0.0068 & 0.0069 & 0.1388 \\
SE & 0.0082 & 0.0086 & 0.0087 & 0.0146 & 0.0063 & 0.0066 & 0.0066 & 0.0076 & 0.0045 & 0.0048 & 0.0048 & 0.0120 \\
Model RMSE &   & 5.3842 & 5.5355 & 6.7473 &   & 7.0478 & 7.1922 & 7.6708 &   & 10.1061 & 10.1970 & 10.5776 \\
RMSE$_l$ &   & 1.4139 & 1.4334 & 1.7227 &   & 1.4301 & 1.4470 & 1.5803 &   & 1.4495 & 1.4568 & 1.5132 \\
RMSE$_m$ &   & 0.9852 & 0.9969 & 1.0153 &   & 0.9974 & 1.0060 & 1.0022 &   & 1.0097 & 1.0135 & 1.0491 \\
\multicolumn{13}{c}{\emph{Panel B: p = 100}}\\
Bias & 0.0011 & 0.0149 & 0.0037 & -0.0545 & 0.0009 & 0.0143 & 0.0080 & -0.0745 & -0.0005 & 0.0153 & 0.0101 & -0.0204 \\
Infeasible bias & -0.0004 & 0.0015 & -0.0127 & -0.0661 & -0.0010 & 0.0004 & -0.0070 & -0.0780 & -0.0003 & -0.0006 & -0.0047 & -0.0255 \\
MC RMSE & 0.0001 & 0.0003 & 0.0001 & 0.0381 & 0.0000 & 0.0003 & 0.0001 & 0.0575 & 0.0000 & 0.0003 & 0.0001 & 0.0142 \\
SD & 0.0084 & 0.0087 & 0.0101 & 0.1883 & 0.0065 & 0.0069 & 0.0074 & 0.2291 & 0.0047 & 0.0055 & 0.0049 & 0.1179 \\
SE & 0.0082 & 0.0085 & 0.0086 & 0.0136 & 0.0063 & 0.0064 & 0.0066 & 0.0122 & 0.0045 & 0.0046 & 0.0046 & 0.0084 \\
Model RMSE &   & 5.3482 & 5.5007 & 7.7637 &   & 6.9908 & 7.1358 & 8.2569 &   & 9.9905 & 10.1178 & 12.5143 \\
RMSE$_l$ &   & 1.3900 & 1.4124 & 1.7716 &   & 1.4082 & 1.4257 & 1.6887 &   & 1.4239 & 1.4334 & 1.8217 \\
RMSE$_m$ &   & 0.9751 & 0.9891 & 1.2031 &   & 0.9889 & 0.9978 & 1.1811 &   & 0.9991 & 1.0040 & 1.0361 \\
\multicolumn{13}{c}{\emph{Panel C: p = 300}}\\
Bias & -0.0004 & 0.0053 & -0.0109 & -0.1452 & 0.0002 & 0.0049 & -0.0047 & -0.1034 & -0.0005 & 0.0061 & -0.0008 & -0.0718 \\
Infeasible bias & 0.0006 & 0.0002 & -0.0157 & -0.1515 & 0.0001 & -0.0002 & -0.0108 & -0.1181 & 0.0002 & 0.0008 & -0.0056 & -0.0822 \\
MC RMSE & 0.0001 & 0.0001 & 0.0002 & 0.0466 & 0.0000 & 0.0001 & 0.0001 & 0.0416 & 0.0000 & 0.0001 & 0.0000 & 0.0233 \\
SD & 0.0083 & 0.0089 & 0.0091 & 0.1605 & 0.0066 & 0.0069 & 0.0064 & 0.1766 & 0.0047 & 0.0040 & 0.0041 & 0.1354 \\
SE & 0.0082 & 0.0084 & 0.0086 & 0.0153 & 0.0063 & 0.0065 & 0.0065 & 0.0128 & 0.0045 & 0.0045 & 0.0046 & 0.0085 \\
Model RMSE &   & 5.3114 & 5.4909 & 6.9683 &   & 6.9550 & 7.1161 & 14.0779 &   & 9.9335 & 10.0883 & 12.2875 \\
RMSE$_l$ &   & 1.3760 & 1.3981 & 1.5737 &   & 1.3945 & 1.4103 & 1.8905 &   & 1.4092 & 1.4186 & 1.5550 \\
RMSE$_m$ &   & 0.9713 & 0.9844 & 1.1450 &   & 0.9841 & 0.9926 & 1.1590 &   & 0.9937 & 0.9984 & 1.0941 \\
\hline
\end{tabular}
\begin{tablenotes}[para,flushleft]
\textbf{Note:} The figures in the table are average values and frequencies over 100 replications by estimation method, i.e.,  IFE by \citet{bai2009panel}, and our panel DML-IFE with Lasso, gradient boosting, and neural network. The true structural parameter $\theta$ is 1.  Standard errors in parenthesis are clustered at the firm level.  Additional details on panel DML-IFE estimation: cross-fitting with 5 folds. Lasso hyperparameter is selected from the model with minimum cross-validated error; gradient boosting and neural network hyperparameters are tuned with random search. 
\end{tablenotes}
\end{threeparttable}}
\end{table}


\begin{table}[t!]
\centering
\caption{Monte Carlo Results: DGP 2, $N = 500$}
\label{tab:dgp2_n500}
\scalebox{.6}{
\begin{threeparttable}
\begin{tabular}{lcccccccccccc}
\vspace{-3mm}\\
\hline\hline
\vspace{-3mm}\\
 & \multicolumn{4}{c}{$T=30$} & \multicolumn{4}{c}{$T=50$} & \multicolumn{4}{c}{$T=100$} \\
\cmidrule(lr){2-5}\cmidrule(lr){6-9}\cmidrule(lr){10-13}
 & IFE & Lasso & Boosting & NNet &IFE & Lasso & Boosting &  NNet &IFE & Lasso & Boosting &  NNet\\
\hline
\multicolumn{13}{c}{\emph{Panel A: p = 50}}\\
Bias & 0.0213 & 0.0790 & 0.0674 & 0.0723 & 0.0216 & 0.0743 & 0.0687 & 0.0710 & 0.0212 & 0.0749 & 0.0662 & 0.0622 \\
Infeasible bias & 0.1589 & 0.0193 & 0.0077 & 0.0031 & 0.1600 & 0.0143 & 0.0055 & 0.0048 & 0.1598 & 0.0085 & 0.0023 & 0.0003 \\
MC RMSE & 0.0005 & 0.0065 & 0.0048 & 0.0056 & 0.0005 & 0.0057 & 0.0049 & 0.0055 & 0.0005 & 0.0057 & 0.0045 & 0.0041 \\
SD & 0.0085 & 0.0177 & 0.0165 & 0.0205 & 0.0063 & 0.0131 & 0.0140 & 0.0213 & 0.0051 & 0.0115 & 0.0119 & 0.0145 \\
SE & 0.0079 & 0.0087 & 0.0087 & 0.0089 & 0.0061 & 0.0068 & 0.0069 & 0.0096 & 0.0043 & 0.0052 & 0.0052 & 0.0053 \\
Model RMSE &   & 5.5399 & 5.6413 & 5.7349 &   & 7.2488 & 7.3185 & 7.4667 &   & 10.3707 & 10.3552 & 10.5421 \\
RMSE$_l$ &   & 1.4995 & 1.5095 & 1.5213 &   & 1.5114 & 1.5182 & 2.1457 &   & 1.5259 & 1.5183 & 1.5267 \\
RMSE$_m$ &   & 1.0246 & 1.0327 & 1.0301 &   & 1.0335 & 1.0386 & 1.0359 &   & 1.0429 & 1.0404 & 1.0418 \\
\multicolumn{13}{c}{\emph{Panel B: p = 100}}\\
Bias & 0.0223 & 0.0512 & 0.0383 & 0.0348 & 0.0223 & 0.0446 & 0.0364 & 0.0422 & 0.0211 & 0.0424 & 0.0352 & 0.0310 \\
Infeasible bias & 0.1572 & 0.0224 & 0.0038 & 0.0019 & 0.1589 & 0.0140 & 0.0034 & 0.0028 & 0.1588 & 0.0080 & 0.0015 & -0.0022 \\
MC RMSE & 0.0006 & 0.0028 & 0.0017 & 0.0015 & 0.0005 & 0.0021 & 0.0014 & 0.0081 & 0.0005 & 0.0018 & 0.0013 & 0.0011 \\
SD & 0.0081 & 0.0148 & 0.0142 & 0.0171 & 0.0062 & 0.0099 & 0.0112 & 0.0798 & 0.0046 & 0.0073 & 0.0075 & 0.0110 \\
SE & 0.0079 & 0.0084 & 0.0085 & 0.0089 & 0.0061 & 0.0065 & 0.0066 & 0.0112 & 0.0043 & 0.0047 & 0.0047 & 0.0050 \\
Model RMSE &   & 5.4753 & 5.5768 & 5.8382 &   & 7.1460 & 7.2085 & 7.4741 &   & 10.1907 & 10.2088 & 10.5915 \\
RMSE$_l$ &   & 1.4552 & 1.4669 & 1.4993 &   & 1.4639 & 1.4686 & 2.3143 &   & 1.4736 & 1.4698 & 1.5077 \\
RMSE$_m$ &   & 1.0059 & 1.0162 & 1.0198 &   & 1.0141 & 1.0184 & 1.0230 &   & 1.0211 & 1.0204 & 1.0240 \\
\multicolumn{13}{c}{\emph{Panel C: p = 300}}\\
Bias & 0.0210 & 0.0303 & 0.0114 & -0.3058 & 0.0220 & 0.0238 & 0.0122 & -0.0917 & 0.0209 & 0.0210 & 0.0113 & -0.0134 \\
Infeasible bias & 0.1613 & 0.0188 & 0.0006 & -0.1630 & 0.1603 & 0.0128 & -0.0006 & -0.1178 & 0.1602 & 0.0101 & -0.0000 & -0.0284 \\
MC RMSE & 0.0005 & 0.0011 & 0.0003 & 0.3307 & 0.0005 & 0.0006 & 0.0002 & 0.0263 & 0.0005 & 0.0005 & 0.0002 & 0.0008 \\
SD & 0.0080 & 0.0119 & 0.0123 & 0.4895 & 0.0064 & 0.0070 & 0.0098 & 0.1343 & 0.0047 & 0.0049 & 0.0059 & 0.0245 \\
SE & 0.0080 & 0.0083 & 0.0084 & 0.0971 & 0.0061 & 0.0064 & 0.0065 & 0.0100 & 0.0043 & 0.0045 & 0.0045 & 0.0058 \\
Model RMSE &   & 5.3991 & 5.5375 & 49.4023 &   & 7.0737 & 7.1547 & 26.0477 &   & 10.0867 & 10.1198 & 11.0700 \\
RMSE$_l$ &   & 1.4240 & 1.4336 & 15.6368 &   & 1.4329 & 1.4361 & 3.2054 &   & 1.4409 & 1.4353 & 1.4955 \\
RMSE$_m$ &   & 0.9954 & 1.0037 & 6.0486 &   & 1.0019 & 1.0065 & 1.7621 &   & 1.0080 & 1.0069 & 1.0330 \\
\hline
\end{tabular}
\begin{tablenotes}[para,flushleft]
\textbf{Note:} The figures in the table are average values and frequencies over 100 replications by estimation method, i.e.,  IFE by \citet{bai2009panel}, and our panel DML-IFE with Lasso, gradient boosting, and neural network. The true structural parameter $\theta$ is 1.  Standard errors in parenthesis are clustered at the firm level.  Additional details on panel DML-IFE estimation: cross-fitting with 5 folds. Lasso hyperparameter is selected from the model with minimum cross-validated error; gradient boosting and neural network hyperparameters are tuned with random search. 
\end{tablenotes}
\end{threeparttable}}
\end{table}


\begin{table}[t!]
\centering
\caption{Monte Carlo Results: DGP 3, $N = 500$}
\label{tab:dgp3_n500}
\scalebox{.6}{
\begin{threeparttable}
\begin{tabular}{lcccccccccccc}
\vspace{-3mm}\\
\hline\hline
\vspace{-3mm}\\
 & \multicolumn{4}{c}{$T=30$} & \multicolumn{4}{c}{$T=50$} & \multicolumn{4}{c}{$T=100$} \\
\cmidrule(lr){2-5}\cmidrule(lr){6-9}\cmidrule(lr){10-13}
 & IFE & Lasso & Boosting & NNet &IFE & Lasso & Boosting &  NNet &IFE & Lasso & Boosting &  NNet\\
\hline
\multicolumn{13}{c}{\emph{Panel A: p = 50}}\\
Bias & 0.1980 & 0.1158 & 0.0844 & 0.1052 & 0.1996 & 0.1049 & 0.0683 & 0.0725 & 0.1989 & 0.0915 & 0.0716 & 0.0839 \\
Infeasible bias & 0.2048 & 0.0411 & 0.0048 & 0.0332 & 0.2046 & 0.0283 & -0.0049 & -0.0197 & 0.2049 & 0.0140 & -0.0043 & -0.0158 \\
MC RMSE & 0.0393 & 0.0138 & 0.0078 & 0.0115 & 0.0399 & 0.0112 & 0.0049 & 0.0185 & 0.0396 & 0.0085 & 0.0053 & 0.0078 \\
SD & 0.0090 & 0.0190 & 0.0265 & 0.0212 & 0.0065 & 0.0146 & 0.0155 & 0.1156 & 0.0048 & 0.0131 & 0.0137 & 0.0272 \\
SE & 0.0085 & 0.0089 & 0.0092 & 0.0098 & 0.0066 & 0.0071 & 0.0073 & 0.0101 & 0.0047 & 0.0054 & 0.0055 & 0.0077 \\
Model RMSE &   & 5.6204 & 6.1291 & 5.8645 &   & 7.3037 & 7.8751 & 7.7826 &   & 10.3815 & 10.8529 & 11.1051 \\
RMSE$_l$ &   & 1.5427 & 1.6147 & 1.6337 &   & 1.5449 & 1.5974 & 1.6108 &   & 1.5406 & 1.5746 & 1.9806 \\
RMSE$_m$ &   & 1.0327 & 1.0733 & 1.0446 &   & 1.0394 & 1.0717 & 1.0732 &   & 1.0426 & 1.0632 & 1.0517 \\
\multicolumn{13}{c}{\emph{Panel B: p = 100}}\\
Bias & 0.1993 & 0.0786 & 0.0467 & 0.0611 & 0.1990 & 0.0675 & 0.0299 & 0.0488 & 0.1987 & 0.0568 & 0.0329 & 0.0387 \\
Infeasible bias & 0.2052 & 0.0405 & 0.0060 & 0.0232 & 0.2056 & 0.0274 & -0.0074 & -0.0114 & 0.2043 & 0.0151 & -0.0067 & -0.0185 \\
MC RMSE & 0.0398 & 0.0063 & 0.0025 & 0.0043 & 0.0397 & 0.0046 & 0.0010 & 0.0027 & 0.0395 & 0.0033 & 0.0011 & 0.0019 \\
SD & 0.0088 & 0.0101 & 0.0172 & 0.0230 & 0.0069 & 0.0086 & 0.0100 & 0.0174 & 0.0047 & 0.0075 & 0.0077 & 0.0191 \\
SE & 0.0085 & 0.0086 & 0.0091 & 0.0095 & 0.0066 & 0.0067 & 0.0070 & 0.0075 & 0.0047 & 0.0048 & 0.0049 & 0.0055 \\
Model RMSE &   & 5.5034 & 6.0945 & 5.9145 &   & 7.1684 & 7.7863 & 7.6582 &   & 10.2086 & 10.7261 & 10.7949 \\
RMSE$_l$ &   & 1.4834 & 1.5703 & 1.5470 &   & 1.4862 & 1.5473 & 1.5376 &   & 1.4852 & 1.5210 & 1.5659 \\
RMSE$_m$ &   & 1.0103 & 1.0581 & 1.0353 &   & 1.0172 & 1.0537 & 1.0363 &   & 1.0215 & 1.0431 & 1.0345 \\
\multicolumn{13}{c}{\emph{Panel C: p = 300}}\\
Bias & 0.1983 & 0.0550 & 0.0260 & -0.1066 & 0.1992 & 0.0411 & 0.0011 & -0.0126 & 0.1991 & 0.0295 & 0.0034 & -0.0093 \\
Infeasible bias & 0.2070 & 0.0419 & 0.0090 & -0.1038 & 0.2056 & 0.0272 & -0.0130 & -0.0436 & 0.2052 & 0.0154 & -0.0095 & -0.0286 \\
MC RMSE & 0.0394 & 0.0031 & 0.0010 & 0.0456 & 0.0397 & 0.0017 & 0.0001 & 0.0015 & 0.0397 & 0.0009 & 0.0000 & 0.0009 \\
SD & 0.0097 & 0.0089 & 0.0171 & 0.1861 & 0.0072 & 0.0065 & 0.0088 & 0.0366 & 0.0052 & 0.0041 & 0.0049 & 0.0289 \\
SE & 0.0085 & 0.0085 & 0.0091 & 0.0121 & 0.0066 & 0.0065 & 0.0069 & 0.0089 & 0.0047 & 0.0045 & 0.0047 & 0.0066 \\
Model RMSE &   & 5.4626 & 6.0999 & 7.3159 &   & 7.0922 & 7.8025 & 8.3645 &   & 10.0586 & 10.6810 & 11.6587 \\
RMSE$_l$ &   & 1.4488 & 1.5499 & 1.6979 &   & 1.4469 & 1.5197 & 1.5893 &   & 1.4436 & 1.4881 & 1.5466 \\
RMSE$_m$ &   & 0.9971 & 1.0503 & 4.0765 &   & 1.0025 & 1.0451 & 1.0638 &   & 1.0061 & 1.0317 & 1.0444 \\
\hline
\end{tabular}
\begin{tablenotes}[para,flushleft]
\textbf{Note:} The figures in the table are average values and frequencies over 100 replications by estimation method, i.e.,  IFE by \citet{bai2009panel}, and our panel DML-IFE with Lasso, gradient boosting, and neural network. The true structural parameter $\theta$ is 1.  Standard errors in parenthesis are clustered at the firm level.  Additional details on panel DML-IFE estimation: cross-fitting with 5 folds. Lasso hyperparameter is selected from the model with minimum cross-validated error; gradient boosting and neural network hyperparameters are tuned with random search. 
\end{tablenotes}
\end{threeparttable}}
\end{table}

\end{document}